\documentclass[12pt]{article}

\usepackage{scicite}

\usepackage{times}

\usepackage{amsmath, amsfonts, amssymb, mathrsfs}
\usepackage{amssymb}
\usepackage{txfonts}
\usepackage{graphicx}
\usepackage{epsfig}
\usepackage{enumerate}
\usepackage{caption}
\usepackage{subcaption}
\usepackage{ytableau}
\usepackage{mathtools} 
\usepackage{extarrows} 
\usepackage[normalem]{ulem}
\usepackage{comment}

\usepackage{soul}

\newcommand{\be}{\begin{equation}}

\newcommand{\ee}{\end{equation}}

\newenvironment{sciabstract}{%
\begin{quote} \bf}
{\end{quote}}

\title{A Non-Unitary Conformal Field Theory Approach to Two-Dimensional Turbulence}

\author
{Jun Nian,$^{1}$ Xiaoquan Yu,$^{2,3}$ Jinwu Ye$^{4,5}$\\
\\
\normalsize{$^{1}$ International Centre for Theoretical Physics Asia-Pacific, }\\
\normalsize{University of Chinese Academy of Sciences, 100190 Beijing, China}\\
\normalsize{$^{2}$ Graduate School of China Academy of Engineering Physics, Beijing 100193, China}\\
\normalsize{$^3$ Department of Physics, Centre for Quantum Science, }\\
\normalsize{and Dodd-Walls Centre for Photonic and Quantum Technologies, }\\
\normalsize{University of Otago, Dunedin, New Zealand}\\
\normalsize{$^4$ Institute for Theoretical Sciences,  Westlake University,  Hangzhou,  310024,  Zhejiang,  China}\\
\normalsize{$^5$ Department of Physics and Astronomy,  Mississippi State University,  MS,  39762,  USA}\\
\\
}

\date{}

\begin{document} 

% Double-space the manuscript.

\baselineskip24pt

% Make the title.

\maketitle

% Place your abstract within the special {sciabstract} environment.

\vspace{-4mm}
\begin{sciabstract}
{Fluid turbulence is a far-from-equilibrium phenomenon and remains one of the most challenging problems in physics.  Two-dimensional,  fully developed turbulence may possess the largest possible symmetry,  the conformal symmetry.  We focus on the steady-state solution of two-dimensional bounded turbulent flow and propose a $c=0$ boundary logarithmic conformal field theory for the inverse energy cascade and another bulk conformal field theory in the classical limit $c\rightarrow -\infty$ for the direct enstrophy cascade.  We show that these theories give rise to the Kraichnan-Batchelor scaling $k^{-3}$ and the Kolmogorov-Kraichnan scaling $k^{-5/3}$ for the enstrophy and the energy cascades, respectively,  with the expected cascade directions,  fluxes,  and fractal dimensions.  We also made some new predictions for future numerical simulations and experiments to test.}

\end{sciabstract}

% In setting up this template for *Science* papers, we've used both
% the \section* command and the \paragraph* command for topical
% divisions.  Which you use will of course depend on the type of paper
% you're writing.  Review Articles tend to have displayed headings, for
% which \section* is more appropriate; Research Articles, when they have
% formal topical divisions at all, tend to signal them with bold text
% that runs into the paragraph, for which \paragraph* is the right
% choice.  Either way, use the asterisk (*) modifier, as shown, to
% suppress numbering.

\section*{Introduction}

Fluid turbulence happens at scales ranging from a water tab and a geographical storm to the size of a galaxy. Although the underlying hydrodynamic equation, i.e., the Navier-Stokes equation, has been known for over a century,  many fundamental questions on turbulence remain unresolved.  In the fully developed homogeneous turbulence, there is a scale range called the inertial range, in which the statistically stationary turbulence forms a self-similar steady flow.  Kolmogorov first found that assuming a constant energy flux per mass, the three-dimensional (3d) turbulence in the inertial range has a kinetic energy spectrum obeying the scaling law $E(k)\sim k^{-5/3}$~\cite{Kolmogorov1, Kolmogorov2}.  Unlike the 3d case \cite{TDLee},  the two-dimensional (2d) turbulence conserves both energy and enstrophy in the inviscid limit.  Due to these two conservation laws,  Kraichnan proposed two cascades in the inertial range for 2d turbulence \cite{Kraichnan}: one with constant energy flux and kinetic energy spectrum $k^{-5/3}$ extending from the intermediate injection scale toward the largest scale available; the other with constant enstrophy flux and kinetic energy spectrum $k^{-3}$ extending from the injection scale down to the viscous scale, known as the inverse energy cascade and the direct enstrophy cascade, respectively. These scaling laws have been extensively tested in numerical simulations \cite{Chen2003,  numerics1,  numerics2,  Falkovich2007,  Chen2009,  Boffeta2020},  suggesting a possible underlying conformal symmetry.

Conformal field theory (CFT) is a powerful tool for understanding (1+1)d quantum and 2d classical critical phenomena. Since the turbulence is a driven-dissipative system,  the conventional unitary CFT is unsuitable for describing its two scaling laws in the inertial range.  So,  finding a non-unitary CFT description for the turbulence in the inertial range is desirable.  This program was initiated by Polyakov~\cite{Polyakov:1992yw, Polyakov:1992er}.  However,  despite many numerical attempts \cite{Chen2003,  numerics1,  numerics2,  Falkovich2007,  Chen2009,  Boffeta2020},  none of the CFT approaches \cite{Matsuo:1992zp,  Falkovich:1992yh,  Lowe:1992ym,  Cateau:1993si,  Falkovich:1993fg,  Falkovich2010} reproduced the precise scalings ($k^{-5/3}$ and $k^{-3}$) and the directions of two cascades simultaneously.  There is also an interesting approach using the AdS/CFT correspondence \cite{Adams:2012pj,  Adams:2013vsa},  but it is based on the assumption that the boundary turbulence is described by a unitary CFT.  The fact that turbulence is a driven-dissipative system,  therefore non-unitary,  makes this approach questionable.

In this work, we resolve this long-standing problem of finding the correct non-unitary CFT description for the 2d turbulence in the inertial range,  which produces the precise scalings $k^{-5/3}$ and $k^{-3}$,  the directions of two cascades and also the fractal dimensions in a unified picture.  We also establish some intrinsic relations between the three length scales $(L,  \ell,  a)$ in the inertial range in Fig.~\ref{fig:scalings},  which can be tested in future numerical simulations and experiments.

For the direct enstrophy cascade,  we propose a new CFT description,  the semiclassical $W_2$ conformal field theory.  It is a 2d non-unitary CFT with Virasoro algebra and can be obtained from the minimal models $(p',  p)$ by taking the limit $p$ finite and $p' \to \infty$.  We find that it can provide the precise Kraichnan-Batchelor scaling $E(k) \sim k^{-3 + \mathcal{O} (1/c)}$  for the direct enstrophy cascade as the central charge $c\rightarrow -\infty$.  In addition,  this CFT also correctly predicts a constant enstrophy flux,  a vanishing energy flux,  and almost no fractals in the vorticity cluster boundaries.  For the inverse energy cascade,  we find that the corresponding CFT is a $c=0$ boundary logarithmic CFT,  which is a direct sum of the 2d CFTs,  i.e.,  (($Q$=$1$)-Potts model) $\oplus$ ($O$($N$=0) model).  These two coexisting CFTs describe the energy cascade.  Using this coexisting boundary logarithmic CFT description,  we can derive the precise Kolmogorov-Kraichnan scaling $k^{-5/3}$,  a constant energy flux,  a vanishing enstrophy flux,  and also the two coexisting fractal dimensions $\frac{7}{4}$ and $\frac{4}{3}$ of the vorticity cluster boundaries,  fully consistent with the previous numerical results \cite{numerics1}.  Finally, we show that the infinite conserved quantities in our non-unitary CFTs on both cascades are in the one-to-one correspondence with those in the classical Korteweg-De Vries (KdV) equation.

\section*{Conformal Field Theory Approach to 2D Turbulence}

Let us first review some aspects of the 2d turbulence.  The Navier-Stokes equation (NSE) is
\be\label{eq:NS 1}
\frac{\partial {\bf u}}{\partial t} + {\bf u} \cdot \nabla {\bf u} = - \frac{1}{\rho} \nabla p + \nu \nabla^2 {\bf u} + \frac{1}{\rho} {\bf f}\, ,
\ee
where ${\bf u}$ is the fluid velocity, $\rho$ is the mass density, $p$ is the pressure,  $\nu$ is the viscosity,  and $\mathbf{f}$ is the stirring force.
The turbulence solution emerges at large Reynolds numbers $1 \ll Re \equiv \rho L |{\bf u}| / \nu$.  Let us denote the inertial range of turbulence by $[k_o,  k_c]$,  where $k_c \sim a^{-1}$ with a small length scale cutoff $a$,  and $k_o \sim L^{-1}$ with $L$ denoting the system size.  On the right-hand side of the NSE,  the term $\nu \nabla^2 {\bf u}$ is more relevant when close to the UV cutoff $k_c$,  while the external stirring force ${\bf f}$ is important near the injection scale $k_i \equiv \ell^{-1}$,  where the energy is fed to the system.  In terms of the vorticity $\omega = \epsilon_{\alpha\beta} \partial_\beta {\bf u}_\alpha$,  the (2+1)d incompressible ($\nabla \cdot {\bf u}=0$) NSE becomes
\be\label{eq:NS 2}
  \dot{\omega} + \epsilon_{\alpha\beta}\, \partial_\alpha \psi\, \partial_\beta \partial^2 \psi = \nu\, \partial^2 \omega + F\, ,
\ee
where $\psi$ denotes the stream function,  $\omega =\partial^2 \psi$,  ${\bf u}_\alpha = \epsilon_{\alpha \beta} \partial_\beta \psi$ and $F = \epsilon_{\alpha\beta} \partial_\beta {\bf f}_\alpha / \rho$ for a constant density $\rho$.  We focus on the NSE of vorticity for most parts of the paper and only occasionally use the NSE of velocity.

\begin{figure}[!htb]
      \begin{center}
        \includegraphics[width=0.62\textwidth]{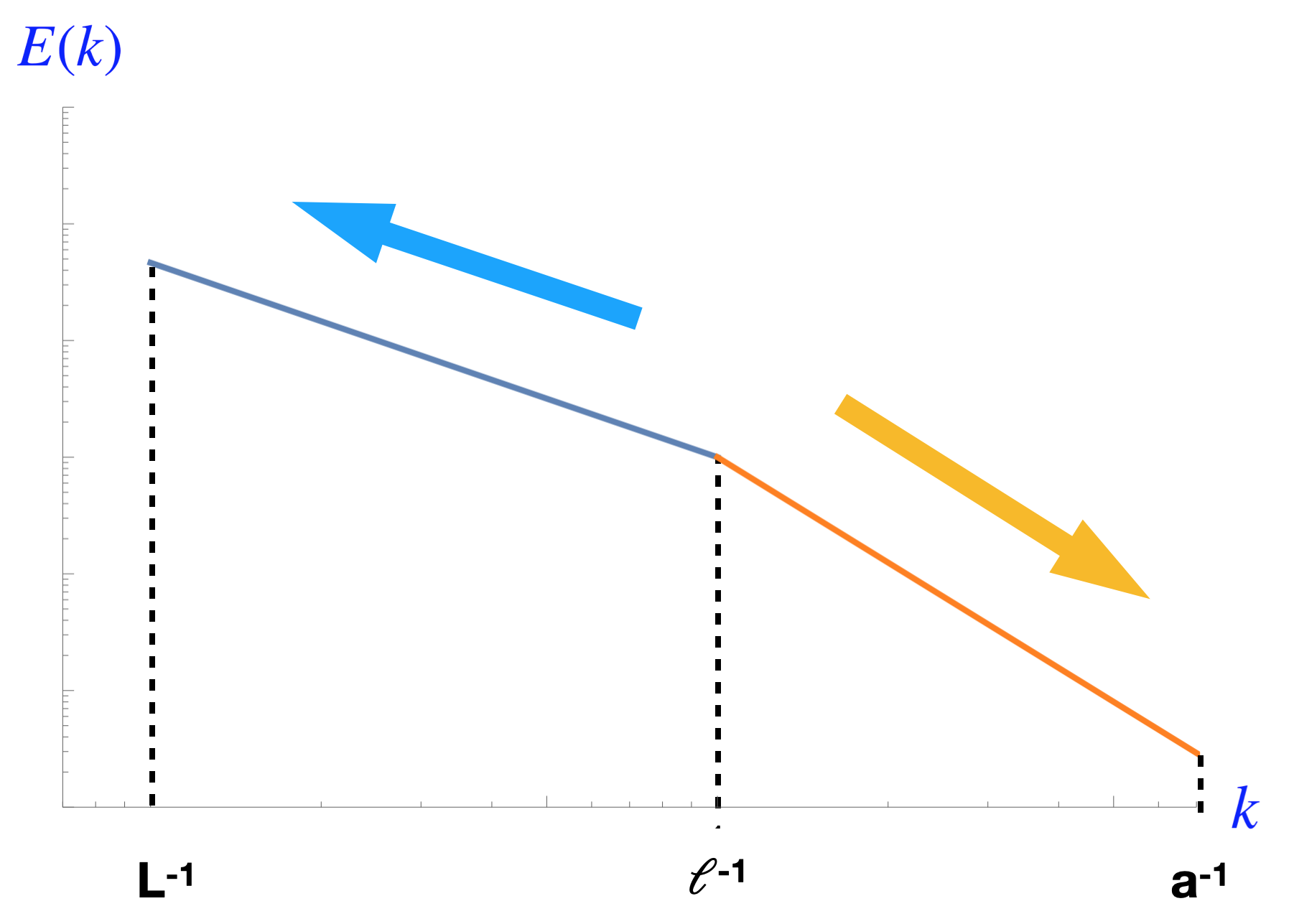}
        \caption{The scalings of 2d turbulence: the blue line has $E(k) \sim k^{-5/3}$, while the yellow line has $E(k) \sim k^{-3}$. The arrows denote the cascade directions. }
        \label{fig:scalings}
      \end{center}
    \end{figure}

As discussed by T.D.~Lee in \cite{Lee}, 2d turbulence cannot support the usual Kolmogorov direct energy cascade of 3d turbulence. Kraichnan proposed a dual cascade picture by considering two different constants of the 2d turbulence, the energy transfer rate $J^{(E)}$ and the enstrophy transfer rate $J^{(H)}$, where the enstrophy is $H \equiv \int d^2 x\, \omega^2$.  By imposing one of the two constants, Kraichnan found that for $k_o \ll k \ll k_i$, the energy transfers at a constant rate $J^{(E)}$ from small scales (large $k$) to large scales (small $k$), which is called the inverse energy cascade, while for $k_i \ll k \ll k_c$ the enstrophy transfers at a constant rate $J^{(H)}$ from large scales (small $k$) to small scales (large $k$), which is called the direct enstrophy cascade. The energy spectrum has the scaling $E(k) \sim k^{-5/3}$ for the inverse energy cascade and $E(k) \sim k^{-3}$ for the direct enstrophy cascade \cite{Kraichnan} (see Fig.~\ref{fig:scalings}).

The simple scalings in the inertial range strongly suggest the existence of an effective conformal symmetry.  Polyakov's pioneer work \cite{Polyakov:1992yw, Polyakov:1992er} opened up the possibility of using 2d conformal field theories to study 2d turbulence.  The main idea is to search for a CFT that captures far-from-equilibrium fluctuations of the vorticity field in the inertial range,  in which the incompressible inviscid ($\nu = 0$) NSE is satisfied as an operator equation,  $\dot{\omega} = - \epsilon_{\alpha\beta}\, \partial_\alpha \psi\, \partial_\beta \partial^2 \psi=0$,  by identifying the fluctuating stream function $\psi$ with a primary field in the CFT.  This is based on the assumption that the system possesses the biggest symmetry in the inertial range,  i.e.,  the conformal symmetry,  a possible emergent symmetry through chaos.  Let $\psi$ be a primary operator with the algebraic fusion rule $\psi \times \psi = \phi + \cdots$,  where $\phi$ is a primary operator with either the highest or the lowest conformal weight depending on the boundary conditions.  Consequently,  $- \epsilon_{\alpha\beta}\, \partial_\alpha \psi\, \partial_\beta \partial^2 \psi$ has a CFT interpretation $\sim |a|^{2 (h_\phi - 2 h_\psi)} \left(L_{-2} \bar{L}_{-1}^2 - \bar{L}_{-2} L_{-1}^2 \right) \phi$,  where $L_{-n}$ and $\bar{L}_{-n}$ are Virasoro generators \cite{DiFrancesco:1997nk}.  If this CFT expression vanishes when $a \to 0$,  the incompressible NSE as an operator equation $\dot{\omega} = 0$ will automatically be satisfied.  This can happen when (i) $h_\phi > 2 h_\psi$; or (ii) $\left(L_{-2} \bar{L}_{-1}^2 - \bar{L}_{-2} L_{-1}^2 \right) \phi$ vanishes or corresponds to a symmetry.  Our proposal in this paper will be the former case.

Hereafter,  we consider a flow in a rectangular region.  The energy $E$ and the enstrophy $H$ can be obtained as
\be
  E = \frac{L^{2 (h_{\widetilde{\alpha}} + h_{\widetilde{\beta}})}}{2} \int d^2 k\, \langle {\bf u} ({\bf k})\cdot {\bf u} (- {\bf k}) \rangle_{\widetilde{\alpha} \widetilde{\beta}}\, ,\quad H = \frac{L^{2 (h_{\widetilde{\alpha}} + h_{\widetilde{\beta}})}}{2} \int d^2 k\, \langle \omega ({\bf k})\, \omega (- {\bf k}) \rangle_{\widetilde{\alpha} \widetilde{\beta}}\, ,
\ee
where $\langle \cdot \rangle_{\widetilde{\alpha}\, \widetilde{\beta}}$ denotes the expectation value of some bulk operators with boundary operators $\widetilde{\alpha}$ and $\widetilde{\beta}$ inserted at two ends of the domain,  ${\bf u} ({\bf k})$ and $\omega ({\bf k})$ are the Fourier transforms of ${\bf u}$ and $\omega$,  and the factors $L^{2 (h_{\widetilde{\alpha}} + h_{\widetilde{\beta}})}$ with the system size $L$ are introduced to compensate the dimensions.  For the case without boundary,  $E$ and $H$ can be defined using the ordinary correlation functions without boundary operator insertions.  The integrands in $E$ and $H$ can be expressed in terms of 2-point functions $\langle \psi\, \psi \rangle$ in the CFT description.

In addition,  we can define the energy flux (or the energy transfer rate) $J^{(E)}$ and the enstrophy flux (or the enstrophy transfer rate) $J^{(H)}$ at a given energy scale $q$:
\begin{align}
\begin{split}
  J^{(E)} (q) & = - L^{2 (h_{\widetilde{\alpha}} + h_{\widetilde{\beta}})} \int_{|{\bf k}| > q} d^2 k\, \langle \dot{\bf u} ({\bf k})\cdot {\bf u} (- {\bf k}) \rangle_{\widetilde{\alpha} \widetilde{\beta}} = L^{2 (h_{\widetilde{\alpha}} + h_{\widetilde{\beta}})} \int_{|{\bf k}| < q} d^2 k\, \langle \dot{\bf u} ({\bf k})\cdot {\bf u} (- {\bf k}) \rangle_{\widetilde{\alpha} \widetilde{\beta}}\, ,\\
  J^{(H)} (q) & = - L^{2 (h_{\widetilde{\alpha}} + h_{\widetilde{\beta}})} \int_{|{\bf k}| > q} d^2 k\, \langle \dot{\omega} ({\bf k})\, \omega (- {\bf k}) \rangle_{\widetilde{\alpha} \widetilde{\beta}} = L^{2 (h_{\widetilde{\alpha}} + h_{\widetilde{\beta}})} \int_{|{\bf k}| < q} d^2 k\, \langle \dot{\omega} ({\bf k})\, \omega (- {\bf k}) \rangle_{\widetilde{\alpha} \widetilde{\beta}}\, ,
\end{split}
\end{align}
where boundary operators have been inserted.  For the case without boundary,  the fluxes can be defined using the ordinary correlation functions without boundary operator insertions.  The integrands can be expressed in terms of 3-point functions $\langle \psi\, \psi\, \psi \rangle$ in the CFT description.

In order to have a constant energy flux or a constant enstrophy flux,  we distinguish two cases: (i) the non-identity operators in the CFT have vanishing vacuum expectation values (VEVs); (ii) all the operators in the CFT have nonvanishing vacuum expectation values.  The former case corresponds to a CFT without boundary,  while the latter case naturally corresponds to a CFT defined in a domain with boundary.  A careful dimensional analysis shows that the energy density $E(k)$ defined by $E = \int^{\infty}_0 dk\, E(k)$ scales differently for the two cases: (i) $E(k) \sim k^{4 h_\psi + 1}$; (ii) $E(k) \sim k^{4 h_\psi - 2 h_\phi + 1}$.  For the enstrophy cascade,  Polyakov has proposed the $(21, 2)$ minimal model as the CFT solution with vanishing VEVs,  which leads to the scaling $E(k) \sim k^{-25/7}$ \cite{Polyakov:1992yw, Polyakov:1992er}.  Later,  many more examples with similar scalings were constructed \cite{Matsuo:1992zp, Falkovich:1992yh, Lowe:1992ym, Cateau:1993si},  but none of the minimal model CFTs can reproduce precisely the scaling $k^{-3}$ \cite{Falkovich:1993fg}.  For the energy cascade,  it remains open how to derive the precise scaling law $E(k) \sim k^{-5/3}$ from a CFT \cite{Oz:2018yaz}.

Suppose that the fusion of three $\psi$'s is $\psi\times \psi\times \psi = \chi + \cdots$,  where $\chi$ is a primary operator with the lowest or the highest conformal weight in the fusion channels,  depending on the boundary conditions.  By requiring a constant enstrophy or a constant energy flux,  we can uniquely fix the operator $\chi$,  which helps identify the CFT description of 2d turbulence.

\section*{CFT Description of 2D Direct Enstrophy Cascade}

Starting from the first papers on this subject \cite{Polyakov:1992yw, Polyakov:1992er},  many attempts to find the CFT description of 2d turbulence reply on a particular class of 2d conformal field theories,  the minimal models \cite{Matsuo:1992zp, Falkovich:1992yh, Lowe:1992ym, Cateau:1993si}.  However,  none of them can produce the exact Kraichnan-Batchelor scaling $E(k) \sim k^{-3}$ for the enstrophy cascade.

To resolve this problem,  we propose a candidate CFT,  which is not a minimal model but can be obtained as a special limit of the minimal models.  More precisely,  we consider a minimal model $(p',\, p)$,  and take the limit $p' \to \infty$ keeping $p$ finite,  where $p$ and $p'$ are both positive integers.  Let us briefly discuss this construction.

The central charge of the minimal model $(p',\, p)$ is $c = 1 - 6 (p - p')^2 / (p\, p')$,  which becomes $c = - 6 p' / p + \mathcal{O} (1)$ in the limit $p$ finite and $p' \to \infty$,  i.e.,  $c \to - \infty$ in this limit.  The primary operators of a minimal model $(p', \, p)$ are denoted by $\phi_{r,\, s}$ with two positive integers $r$ and $s$ in the range $1 \leq r \leq p' - 1,\, 1 \leq s \leq p - 1$.  The conformal weight of the primary operator $\phi_{r,\, s}$ is $h (r,\, s) = \left[ (r p - s p')^2 - (p - p')^2\right] / (4 p\, p')$.  Using these notations,  we see that the enstrophy cascade of 2d turbulence can be realized by the Kac operators $\psi = \phi_{3,\, 1}$ and $\phi = \phi_{5,\, 1}$,  which satisfy the fusion rule $\phi_{3,1} \times \phi_{3,1} = \phi_{1,1} + \phi_{3,1} + \phi_{5,1} = I + \psi + \phi$,  where $\phi_{5,1}$ has the lowest negative conformal weight in the fusion products.  More precisely,  these operators have the conformal weights $h_\psi = -1 + 2 p / p' + \mathcal{O} \left((p/p')^{-2}\right)$,  $h_\phi = -2 + 6 p / p' + \mathcal{O} \left((p/p')^{-2}\right)$,  and vanishing vacuum expectation values $\langle \psi \rangle = \langle \phi \rangle = 0$.  Finally,  these results lead to the desired scaling
\be
  E(k) \sim k^{4 h_\psi + 1} = k^{-3 + \mathcal{O} (c^{-1})}\, ,
\ee  
and the additional constraint $h_\phi - 2 h_\psi = 2 p / p' +  \mathcal{O} \left((p')^{-2}\right) > 0$ is also satisfied.  It is interesting to observe that with a large but finite central charge $c$, the infrared divergence of the enstrophy spectrum is avoided.  Therefore,  in the absence of external forces, the incompressible NSE is solved as a CFT operator equation in the strict inviscid limit $\nu = 0$.

To have a nonvanishing enstrophy flux $J^{(H)}$,  we have to keep a small but nonzero viscosity $\nu$ in the NSE of vorticity.  Then,  with the small length scale cutoff $a$, we can compute the enstrophy flux
\be
  J^{(H)} (q) = - \int_{q < |{\bf k}| < a^{-1}} d^2 k\, \langle \dot{\omega} ({\bf k})\, \omega (- {\bf k}) \rangle \sim - \frac{\nu \left((a^{-1})^{6 + 4\, h_\psi} - q^{6 + 4\, h_\psi} \right)}{6 + 4\, h_\psi} \approx - \frac{\nu}{2\, a^2}\, ,
\ee
where the probing energy scale $q \ll a^{-1}$,  and we have used $h_\psi = -1$.  Hence,  a constant enstrophy flux $J^{(H)}$ implies that $\nu \propto a^2$,  which also holds in the definition of the 2d Kraichnan scale \cite{Bartuccelli1993}.  Keeping this relation,  we can again push both $\nu$ and $a$ to zero,  still having a nonvanishing $J^{(H)}$.  Because the sign of $J^{(H)}$ is negative,  the direction of the enstrophy cascade is from small $k$ to large $k$,  i.e.,  the enstrophy cascade is direct.  Following similar steps,  we can keep a small but nonvanishing viscosity $\nu$ in the NSE of velocity and compute the energy flux $J^{(E)}$ for the enstrophy cascade,  and the result for $h_\psi = -1$ is
\be
  J^{(E)} (q) = - \int_{q < |{\bf k}| < a^{-1}} d^2 k\, \langle \dot{\bf u} ({\bf k})\cdot {\bf u} (- {\bf k}) \rangle \sim - \nu\, \textrm{log}\, \frac{1}{q a} = - \nu\, \textrm{log} \frac{1}{q \sqrt{\nu}} + \mathcal{O} (\nu)\, ,
\ee
which is zero in the limit $\nu \to 0$ and $q\, a \to 0$.  In the absence of boundary,  the term $- (\nabla p) / \rho$ in the incompressible NSE of velocity leads to integrating a total derivative in $J^{(E)}$ and hence does not contribute.  Therefore,  the energy flux $J^{(E)}$ vanishes in the direct enstrophy cascade,  as expected.

The CFT approach treats only two spatial dimensions.  If we take into account the time-dependence in the vorticity $\omega$ and the velocity ${\bf u}$,  the fluxes $J^{(H)}$ and $J^{(E)}$ have dimensions the same as in the old arguments.  Hence,  a simple dimensional analysis shows a relation for the 2d enstrophy cascade,  $E(k) = C'\cdot (J^{(H)})^{2/3}\cdot k^{-3}$, the same as Kraichnan's result \cite{Kraichnan}.  However,  our CFT approach explicitly predicts the $\nu$- and $a$-dependencies in $J^{(H)}$.

The solution above suggests not just one CFT,  but a class of CFTs with different values of finite $p$.  The union of this class of CFTs has a more elegant description as a special coset model,  the semiclassical $W_2$ CFT \cite{Gaberdiel:2010pz, Perlmutter:2012ds} defined as the coset Wess-Zumino-Witten (WZW) model $\left(\mathfrak{su} (N)_k \oplus \mathfrak{su} (N)_1 \right) / \mathfrak{su} (N)_{k+1}$ in the semiclassical limit $N=2$ and $k \to -3^{-}$,  which is discussed in the supporting materials.

Although the numerical work \cite{numerics2} suggests that in the direct enstrophy cascade, the conformal invariance is mildly broken by the multifractality \cite{Sreenivasan},  the generalized fractal dimensions $D_q \approx 1$ for large $q$ are valid to a large extent in $k_i \ll k \ll k_c \sim a^{-1}$.  Because the fractal dimension and the diffusivity obey $D = 1 + \kappa / 8$,  a vanishing diffusivity $\kappa = 0$ implies almost no fractals.  According to the relation $c = (8 - 3 \kappa) (\kappa - 6) / (2 \kappa)$ (see \cite{Falkovich2007}),  this is consistent with our proposal of the CFT in the classical limit $c \to - \infty$.

\section*{Boundary CFT Description of 2D Inverse Energy Cascade}

The numerical work \cite{numerics1} suggested that there might be a logarithmic CFT with $c = 0$ governing the inverse energy cascade of the 2d turbulence.  This class of CFT has been reviewed in \cite{Cardy:2013rqg}.  We find that for the inverse energy cascade, we need a boundary logarithmic CFT as a direct sum of two constituting 2d CFTs,  (($Q$=$1$)-Potts model) $\oplus$ ($O$($N$=0) model),  both of which account for the Kolmogorov-Kraichnan scaling $k^{-5/3}$ and two coexisting fractal dimensions $\frac{7}{4}$ and $\frac{4}{3}$.  In this picture, it is crucial to insert some boundary operators \cite{QuantumImpurityProblem},  which have precise physical meanings in the constituting boundary $Q$-Potts model and, at the same time, constrain the operator content of the direct sum CFT in the bulk.  In this boundary logarithmic CFT,  the bulk operators have nonzero VEVs.  

% This is consistent with the statements in \cite{Oz:2018yaz} that in the absence of VEVs, the 2d turbulence can only have scale invariance instead of conformal symmetry except for the direct enstrophy cascade.  In contrast, in the presence of VEVs, the conformal symmetry may exist but is spontaneously broken.

\begin{figure}[htb!]
\begin{center}
  \includegraphics[width=0.61\textwidth]{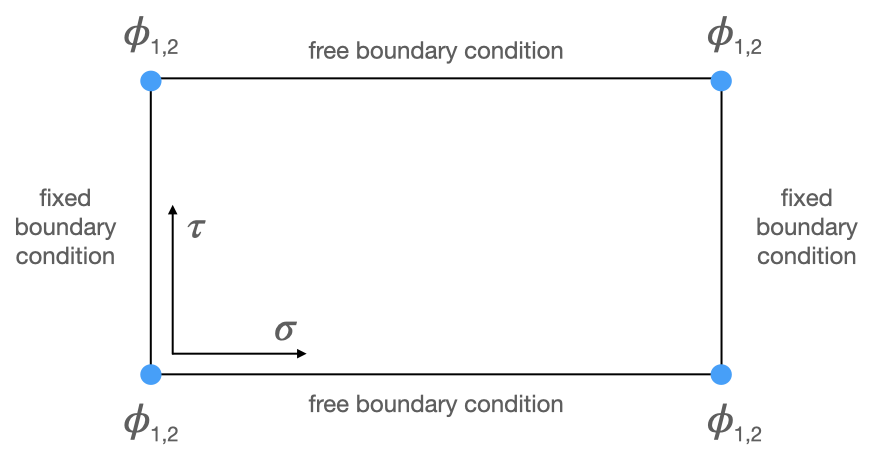}
  \caption{The 2d turbulence in a rectangular region with a boundary}\label{fig:rectangle}
\end{center}
\end{figure}

\begin{figure}[htb!]
\begin{center}
  \includegraphics[width=0.52\textwidth]{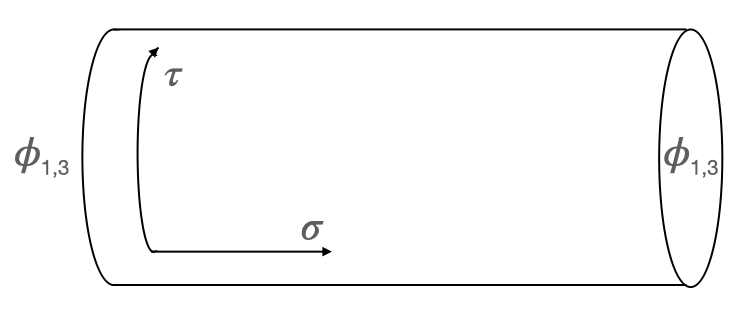}
  \caption{The cylinder region with boundary operators at two ends}\label{fig:cylinder}
\end{center}
\end{figure}

For simplicity,  let us consider a 2d rectangular region with the coordinates $(\tau,\, \sigma)$ (see Fig.~\ref{fig:rectangle}).
Similar to Cardy's treatment of 2d percolation in \cite{Cardy:1991cm},  we impose the fixed boundary condition on the two opposite sides and the free boundary condition on the other two opposite sides by inserting four Kac operators $\phi_{1,2}$ at four corners of the rectangular region.  The reason is that because of the fusion rule $\phi_{1,2} \times \phi_{1,2} = I + \phi_{1,3}$,  the two boundary operators $I$ and $\phi_{1,3}$ correspond to the free and the fixed boundary conditions respectively.  By gluing two boundaries with the same free boundary condition,  we can also transform the region into a cylinder with two boundary operators $\phi_{1,3}$ inserted at two ends (see Fig.~\ref{fig:cylinder}).  In a boundary CFT,  all the operators should have nonvanishing vacuum expectation values.  When two boundary operators $\widetilde{\alpha}$ and $\widetilde{\beta}$ are inserted at two ends of a cylinder,  only the operators $i$ with nonzero fusion coefficients $\mathcal{N}_{\widetilde{\alpha} \widetilde{\beta}}^i$ can appear in the bulk.  In our construction,  two boundary operators are $\phi_{1,3}$ in the constituting ($Q$=1)-Potts model,  and their fusion rule is $\phi_{1,3} \times \phi_{1,3} = I + \phi_{1,3} + \phi_{1,5}$.  Hence,  in this case, only the Kac operators $I$,  $\phi_{1,3}$ and $\phi_{1,5}$ can appear in the bulk.  To establish a CFT description of the 2d inverse energy cascade,  we need to identify the bulk operators $\psi$,  $\phi$, and $\chi$. 

The bulk direct sum CFT,  (($Q$=1)-Potts model) $\oplus$ ($O$($N$=0) model),  has a spectrum from both constituting CFTs,  and the bulk energy operator in the $O(N)$ model is $\phi_{1,3}$ \cite{Dotsenko:1984nm},  which is present in the allowed bulk operator content.  Hence,  we identify $\psi$ as the bulk energy operator,  i.e.,  $\psi = \phi_{1,3}$.  In the limit $c \to 0$,  the conformal weights of the operators $\phi_{r,s}$ in the Kac table are $h_{r,  s} (c) = \left[(3 r - 2 s)^2 - 1 \right] / 24 + \mathcal{O} (c)$.  Therefore,  the operator $\psi = \phi_{1,3}$ in the limit $c \to 0$ has the conformal weights $h_\psi = \bar{h}_\psi = \frac{1}{3}$.

A candidate for the bulk operator $\phi$ is $\phi_{1,5}$ from the constituting $O$($N$=0) model,  which has the conformal weights $(h, \bar{h})=(2,2)$.  With this choice, we have $h_\phi > 2 h_\psi$,  and then in the absence of external forces, the inviscid NSE as a CFT operator equation will be solved.  However,  there is a subtlety of logarithmic CFT.  As Cardy shows in \cite{Cardy:2013rqg} when the conformal weights of two different operators coincide in the limit $c \to 0$,  they form a logarithmic pair, and one can define a new operator with nonvanishing correlation functions from this pair.  The operator $T \overline{T}$ always shows up in the bulk operator product expansion (OPE) as a descendant of the vacuum operator $I$,  and it has the same conformal weights as $\phi_{1, 5}$ in the limit $c \to 0$.  Hence,  they form a logarithmic pair.  Instead of just $\phi_{1,5}$,  we make the identification $\phi = C_\phi\cdot \phi_{1,5} + T \overline{T} / \delta$,  with two constants $C_\phi$ and $\delta$ of order $\mathcal{O} (c^0)$ and $\mathcal{O} (c)$ respectively (see supporting materials).

Knowing the operators $\psi$ and $\phi$ in the CFT description for the energy cascade,  we can compute the energy scaling.  The total energy is $E = L^{2 (h_{\widetilde{\psi}} + h_{\widetilde{\psi}})} \int d^2 k\, \langle {\bf u} ({\bf k})\, {\bf u} (- {\bf k}) \rangle_{\widetilde{\psi}\, \widetilde{\psi}} = \int_{L^{-1}}^{\ell^{-1}} dk\, E(k)$,  with $\ell = k_i^{-1} \ll L$ denoting the injection scale.  Hence,  we can read off from the integrand of $E$ that $E(k) = L^{4 h_{\widetilde{\psi}}}\, k\, \langle{\bf u} ({\bf k})\, {\bf u} (- {\bf k}) \rangle_{\widetilde{\psi}\, \widetilde{\psi}} \sim L^{4 h_{\widetilde{\psi}}}\, k^{4 h_\psi - 2 h_\phi + 1}\cdot \langle \phi \rangle_{\widetilde{\psi}\, \widetilde{\psi}}$,  where $\langle \phi \rangle_{\widetilde{\psi}\, \widetilde{\psi}}$ is independent of $k$.  Consequently,  we obtain the scaling for $h_\psi = \frac{1}{3}$ and $h_\phi = 2$
\be
  E(k) \sim k^{4 h_\psi - 2 h_\phi + 1} = k^{-\frac{5}{3}}\, .
\ee
This is exactly the well-known Kolmogorov-Kraichnan scaling.  We can also integrate $E(k)$ to obtain the energy $E \sim - L^{4 h_{\widetilde{\psi}}} \left(L^{- (4 h_\psi - 2 h_\phi + 1)} - \ell^{- (4 h_\psi - 2 h_\phi + 1)} \right) / (4 h_\psi - 2 h_\phi + 1)$.  From this expression we see that for $L \gg l$ the operator $\phi$ with the highest conformal weight contributes the most,  which justifies the identification of $\phi$ as $\phi_{1,5}$ before taking into account its logarithmic partner.

Without introducing a stirring force,  the energy flux $J^{(E)}$ and the enstrophy flux $J^{(H)}$ should vanish,  which can be used to determine the operator $\chi$ in the fusion of $\psi \times \psi \times \psi$.  As mentioned before,  the energy flux $J^{(E)}$ is defined with specific boundary conditions,  not necessarily the free one.  The boundary condition considered here corresponds to two boundary operators $\widetilde{\psi} = \phi_{1,3}$ inserted at each end of the cylinder.  We find the scaling of the energy flux at the IR cutoff $L^{-1}$ is $J^{(E)} (L^{-1}) = - L^{2 (h_{\widetilde{\psi}} + h_{\widetilde{\psi}})} \int_{|{\bf k}| > L^{-1}} d^2 k\, \langle \dot{\bf u} ({\bf k})\, {\bf u} (- {\bf k}) \rangle_{\widetilde{\psi}\, \widetilde{\psi}} \sim - L^{- 2 h_\phi + 2 h_\chi}$,  where $h_{\widetilde{\psi}} = h_\psi$,  and $\chi$ is the operator with the highest conformal weight in the fusion products of $\psi \times \psi \times \psi$.  In the presence of a boundary,  the term $- (\nabla p) / \rho$ in the NSE of velocity leads to a boundary term in $J^{(E)}$; however,  the fixed boundary condition ${\bf u}_\sigma |_{\sigma = 0,\, L} = 0$ eliminates this boundary term.  A constant energy flux requires that $J^{(E)}$ is independent of $L$,  i.e.,  $- 2 h_\phi + 2 h_\chi = 0$.  A similar analysis can be applied to the enstrophy flux $J^{(H)}$,  which leads to the same result.  Hence,  a candidate for the bulk operator $\chi$ is $\phi_{1,5}$ with the conformal weights $(h,\bar{h})=(2, 2)$.  In the constituting $O$($N$=0) model,  the operator $\phi_{1,5}$ and $T \overline{T}$ have the same conformal weights and form a logarithmic pair.  Therefore,  instead of $\phi_{1,5}$,  the operator $\chi$ should be identified as $\chi = C'_\phi\cdot \phi_{1,5} + T \overline{T} / \delta$,  where $C'_\phi$ is a constant of order $\mathcal{O} (c^0)$,  different from the constant $C_\phi$ in $\phi$,  while $\delta$ remains the same.

In order to have a nonvanishing energy flux $J^{(E)}$,  we should introduce an external stirring force ${\bf f}$ near the injection scale $k_i = \ell^{-1}$.  Since the incompressible NSE has already been solved before turning on an external force,  the energy flux $J^{(E)}$ in the presence of ${\bf f}$ becomes
\be
  J^{(E)} (q) = - L^{2 (h_{\widetilde{\psi}} + h_{\widetilde{\psi}})} \int_{q < |{\bf k}| < \ell^{-1}} d^2 k\, \langle \dot{\bf u} ({\bf k})\, {\bf u} (- {\bf k}) \rangle_{\widetilde{\psi} \widetilde{\psi}} = - L^{4 h_{\widetilde{\psi}}} \int_{q < |{\bf k}| < \ell^{-1}} d^2 k\, \langle \frac{1}{\rho}\, \widetilde{\bf f} ({\bf k})\, {\bf u} (- {\bf k}) \rangle_{\widetilde{\psi} \widetilde{\psi}} \, ,\label{eq:energy flux 2}
\ee
where $\widetilde{\bf f} ({\bf k})$ is the external stirring force in the momentum space.  More precisely,  let us consider a stirring force of the form $\widetilde{\bf f} ({\bf k}) = C_f \epsilon_{\alpha\beta} {\bf k}_\beta / (|{\bf k}|^2 + m^2)$,  where $C_f$ is a positive constant controlling the force strength,  and $m$ is the position of $|{\bf k}|$ at the maximum of $\widetilde{\bf f}$,  which can be chosen to be the injection scale $\ell^{-1}$.  Evaluating the energy flux $J^{(E)}$ in the presence of a stirring force,  we find that it has the scaling $J^{(E)} (q) \sim \frac{C_f}{\rho} \left[\left(\frac{L}{\ell} \right)^{2/3} - (L\, q)^{2/3} \right] \approx \frac{C_f}{\rho} \left(\frac{L}{\ell} \right)^{2/3}$.  Requiring that $C_f / \rho \propto a^2$, the energy flux $J^{(E)}$ can take a finite constant value.  The positive sign of $J^{(E)}$ implies that the energy flow direction is from large $k$ to small $k$,  i.e.,  the energy cascade is inverse.  Applying the same method to the enstrophy flux $J^{(H)}$ in the presence of the external stirring force,  we obtain $J^{(H)} \sim \frac{C_f}{\rho} \frac{1}{\ell^2} \left(\frac{L}{\ell} \right)^{2/3} \propto \left(\frac{a}{\ell}\right)^2 \left(\frac{L}{\ell}\right)^{2/3}$.  Hence,  in the strict limit $a / \ell \to 0$,  the enstrophy flux $J^{(H)}$ vanishes for the inverse energy cascade,  which matches the numerical results in the literature (e.g.,  \cite{boffetta_2007}).  An analysis of fluxes in the intermediate range $[\ell^{-1},  q]$ shows that $C_f / \rho \propto \nu$,  which supports the previous requirement $C_f / \rho \propto a^2$.

Taking into account the time-dependence in the vorticity $\omega$ and the velocity ${\bf u}$,  we find that the fluxes $J^{(E)}$ and $J^{(H)}$ have the dimensions expected from dimensional analysis.  Hence,  we rediscover Kraichnan's result for 2d energy cascade \cite{Kraichnan},  $E(k) = C\cdot (J^{(E)})^{2/3}\cdot k^{-5/3}$.  In practice,  the scale $a$ is small but nonvanishing.  Consequently,  $J^{(H)} \ll 1$ sets up a consistency requirement for scales:
\be
  \left(\frac{L}{\ell}\right)^{\frac{2}{3}} \ll \left(\frac{\ell}{a}\right)^2\, ,
\ee
which is obeyed by most experimental data \cite{Kellay2002,  Bruneau2005}.  We expect that a violation of this requirement may indicate the breakdown of the conformal symmetry in the inertial range of 2d turbulence.

\section*{2D Turbulence and KdV Equation}

It is well known that 2d turbulence has infinite conserved quantities.  Now let us discuss these infinite conserved quantities from the CFT point of view.  The main idea is that for a CFT with central charge $c$ there is an inherent integrable quantum KdV equation for the energy-momentum tensor $T \equiv - 2 \pi T_{zz}$ \cite{Bazhanov:1994ft, DiFrancesco:1997nk}:
\be\label{eq:qu KdV}
  \partial_t T = \frac{1}{6} (1 - c)\, \partial_z^3 T - 3\, \partial_z (TT)\, ,
\ee
which has infinite conserved quantities.  For the enstrophy cascade of 2d turbulence,  we have found a CFT description with $c \to - \infty$.  Defining a new function $u \equiv 6 T / c$ and a new time variable $\tau \equiv - c\, t / 6$,  and then taking the limit $c \to - \infty$,  we can prove that the quantum KdV equation becomes the classical KdV equation.  For the energy cascade of 2d turbulence,  our CFT construction is a $c=0$ boundary logarithmic CFT.  In the limit $c \to 0$,  defining a new function $u \equiv - 6 T$ and a new variable $\tau \equiv t / 6$,  we again obtain the standard classical KdV equation.  The classical KdV equation has infinite conserved quantities $I_n \equiv \int_0^{2 \pi} \frac{dt}{2 \pi}\, \left[u^n (z, t) + \cdots \right]$ ($n = 1,\, 2,\, \cdots$) satisfying $\partial_t I_n = 0$ and $\{I_m,\, I_n \} = 0$.  For the 2d turbulence,  the infinite conserved quantities can be expressed in terms of the vorticity $\omega$ as $H_n \equiv \int d^2 x\, \omega^n (x)$.  Hence,  we see that there exists a one-to-one correspondence between the classical conserved quantities $H_n$ in 2d turbulence and $I_n$ in the classical KdV equation from a CFT.

% \begin{comment}

To set up a more precise connection between the conserved quantities of CFT and the ones for turbulence,  it is more convenient to use the Landau-Ginzburg description of the CFT.  However, not all the minimal models currently have Landau-Ginzburg descriptions \cite{DiFrancesco:1997nk}.  In particular,  for the non-unitary minimal models $(p', p)$ with $p$ finite and $p' \to \infty$ the Landau-Ginzburg description is not clear.  Hence, we leave it as a future project for the enstrophy cascade.  For the energy cascade,  fortunately the Landau-Ginzburg descriptions of the $O(N)$ model and the $Q$-Potts model are known,  which provide $T_{zz} = \sum_a : \partial_z \phi_a\, \partial_z \phi_a :$ and $\psi = \sum_a : \phi_a \phi_a :$,  and the vorticity can be expressed as $\omega = \partial^2 \psi = \partial^2 \left(\sum_a : \phi_a \phi_a :\ \right)$.  Hence,  the infinite conserved quantities at the quantum level can be computed in the constituting $O(N)$ model as
\begin{align}
\begin{split}
  \widetilde{H}_n & = \int d^2 z\, \langle \omega^n (z) \rangle =  \int d^2 z\, \left\langle \left[\partial^2 \left( \sum_a : \phi_a^2 :\right) \right]^n \right\rangle \, ,\\
  \widetilde{I}_n & = \int_0^{2 \pi} \frac{dt}{2 \pi}\, \left\langle \left(-6 T(z, t) \right)^n + \cdots \right\rangle = \int_0^{2 \pi} \frac{dt}{2 \pi}\, \left\langle \left(-6 \sum_a : \left(\partial_z \phi_a \right)^2 : \right)^n + \cdots \right\rangle\, .
\end{split}
\end{align}
Both sets of conserved quantities can be expressed as powers of two derivatives in one-to-one correspondence,  but they are not precisely the same.

% \end{comment}

\section*{Discussion}

In this work,  we have found the non-unitary CFT for 2d turbulence with the expected known behaviors.  We also made new sharp predictions to be tested in future numerical simulations and experiments.  We will elaborate on these CFT descriptions in the near future by computing the higher-point correlation functions.  A quantitative understanding of the deviation from exact conformal invariance (e.g., for 4-point correlation functions) will be beneficiary.  Our success at 2d turbulence implies that 3d turbulence may also be studied using non-unitary CFTs,  with the 3d $O$($N$=0) model as a possible candidate.

This work should also have considerable impacts on many other branches in physics and applied mathematics, such as fluid mechanics,  statistical mechanics,  holography (AdS/CFT, dS/(non-unitary) CFT).  The CFT approach to turbulence may open up a new pathway for studying non-unitary CFTs and their gravity duals.  A natural candidate for the correspondence could be the dS/CFT \cite{Strominger:2001pn,  Maldacena:2019cbz} instead of the AdS/CFT.  It was known that the dS geometry has a cosmological horizon even without a black hole,  so our 2d non-unitary CFT solution on the boundary may imply some novel gravity solution in the bulk dS geometry.

As shown in Table~1 of the supporting materials,  we have noticed that the enstrophy cascade of 2d turbulence also corresponds to the dense phase of the 2d $O$($-$2) model with $c = - \infty$,  while the description of the energy cascade includes the dilute phase of the $O$($N$=0) model with $c = 0$.  The relevant fusion rules for $\phi_{r,s}$ in both cascades take the identical form after exchanging $r$ and $s$.  These facts imply some possible hidden duality relation between the two cascades,  which needs to be explored further.

Inspired by the fact that the unitary CFTs with $c \to +\infty$ have quantum chaos \cite{Roberts:2014ifa},  it is intriguing to pursue the exact meaning of quantum chaos for non-unitary theories and find its fingerprint in the $c \to -\infty$ CFTs.  The $c = 0$ logarithmic CFT may have gravity dual \cite{Grumiller:2010rm},  which may also imply some quantum chaos.

% Your references go at the end of the main text, and before the
% figures.  For this document we've used BibTeX, the .bib file
% scibib.bib, and the .bst file Science.bst.  The package scicite.sty
% was included to format the reference numbers according to *Science*
% style.

%BibTeX users: After compilation, comment out the following two lines and paste in
% the generated .bbl file. 

\bibliography{2dConformalTurbulence}

\bibliographystyle{Science}

\section*{Acknowledgments}
We would like to thank Leo A. Pando Zayas,  Cheng Peng,  Yu Tian,  Huajia Wang,  Hongbao Zhang,  and Yunlong Zhang for the discussions.  J. N.  was supported in part by the NSFC under grant No. 12147103.  X.Y. acknowledges the support from NSAF with grant No. U1930403 and NSFC with grant No. 12175215. 

%Here you should list the contents of your Supplementary Materials -- below is an example. 
%You should include a list of Supplementary figures, Tables, and any references that appear only in the SM. 
%Note that the reference numbering continues from the main text to the SM.
% In the example below, Refs. 4-10 were cited only in the SM.     

\section*{Appendix A: Semiclassical $W_2$ CFT and Enstrophy Cascade}\label{app:W2CFT}

In this appendix, we discuss some properties of the spectrum of the $W_2$ CFT in the semiclassical limit $c \to - \infty$,  which can be viewed as an analytic continuation of the $c \to \infty$ case discussed in \cite{Perlmutter:2012ds}.  Before discussing the details of the semiclassical $W_2$ CFT,  let us briefly recall some properties of the more general $W_N$ CFTs~\cite{WCFT}. The $W_N$ CFTs are defined as the coset Wess-Zumino-Witten (WZW) models
\be\label{eq:coset}
  \frac{\mathfrak{su} (N)_k \oplus \mathfrak{su} (N)_1}{\mathfrak{su} (N)_{k+1}}\, ,
\ee
where $\mathfrak{su} (N)_k$ denotes the WZW model with the $\mathfrak{su} (N)$ Kac-Moody algebra at the level $k$.  These theories have higher spin conserved currents, and are conjectured to be dual to AdS$_3$ higher spin gravities \cite{Gaberdiel:2010pz}. The $W_N$ CFT at the level $k$ has the central charge
\be\label{eq:WNCentralCharge}
  c = (N - 1) \left[1 - \frac{N (N+1)}{(N+k) (N+k+1)} \right]\, ,
\ee
and its primary operators can be labeled by $(\Lambda_+,\, \Lambda_-)$, where $\Lambda_+$ and $\Lambda_-$ denote the representations of $\mathfrak{su} (N)_k$ and $\mathfrak{su} (N)_{k+1}$ respectively.  A generic $W_N$ CFT has the $W_N$ algebra instead of the Virasoro algebra.  Only for $N=2$ the corresponding $W$-algebra is the $W_2$ algebra,  i.e.,  the Virasoro algebra~\cite{WCFT,Pope:1991}.

There are two limits of the $W_N$ CFT discussed in the literature. One is the 't Hooft limit
\be
  N \to \infty,\quad k \to \infty,\quad \lambda \equiv \frac{N}{N + k}\,\, \text{finite}\, ,
\ee
which makes the central charge $c \to \infty$ while keeping the $W_N$ CFT unitary.  A different semiclassical limit of the $W_N$ CFTs was discussed in \cite{Gaberdiel:2010pz, Perlmutter:2012ds}, $k \to - N - 1$ and $N \text{ finite}$, which also makes the central charge $c$ large, but the $W_N$ CFT becomes non-unitary. This limit is of particular interest for the 2d turbulence.  Here, we carefully choose the semiclassical limit to be
\be
  k \to (- N - 1)^-,\quad N\,\, \text{finite}\, ,
\ee
i.e., $k$ approches $- N - 1$ from the negative side. In this limit, the central charge becomes large negative $c \to - \infty$. The CFTs with large negative central charges have been discussed previously in the literature, e.g. \cite{DiFrancesco:1997nk}, which usually correspond to the deep semiclassical regime.

The conformal weight of a primary operator $(\Lambda_+,\, \Lambda_-)$,  denoted by $h (\Lambda_+,\, \Lambda_-)$, can be computed as follows \cite{Perlmutter:2012ds}:
\be\label{eq:WNConformalWeight}
  h (\Lambda_+,\, \Lambda_-) = \frac{\Big[(N + k)\, (r_+ - r_-) + r_+ + 1 \Big]^2 - 1}{4 (N + k)\, (N + k + 1)}\, ,
\ee
where $r_\pm$ denote the number of boxes in the representations $\Lambda_\pm$ respectively.  For a $W_2$ CFT in the semiclassical limit,  if we keep the central charge $c$ large but finite and expand $h (\Lambda_+,\, \Lambda_-)$ in $c$, the leading terms will be
\be
  h (r_+, r_-) = - \frac{r_- (r_- + 2)}{24}\, c + \frac{13 r_-^2 + (14 - 12 r_+) r_- - 12 r_+}{24} - \frac{3 (r_+ - r_-) (r_+ + r_- + 2)}{2 c} + \mathcal{O} \left(c^{-1} \right)\, ,
\ee
where $r_\pm$ denote the number of boxes in the Young tableaux of the representations $\Lambda_\pm$. Hence, we can deduce some properties of the spectrum from this formula.  The semiclassical $W_2$ CFT with a large positive central charge $c > 0$ has the spectrum with non-positive conformal weights. More precisely, only the identity operator $(\bullet,\, \bullet)$ has the conformal weight zero, while all the other primary operators have negative conformal weights. For the semiclassical $W_2$ CFT with a large negative central charge $c < 0$, except $(\bullet,\, \bullet)$ all the other primary operators of the form $(\Lambda_+,\, \bullet)$ have negative conformal weights at the order $\mathcal{O} (c^0)$, while all the primary operators $(\Lambda_+,\, \Lambda_-)$ with $\Lambda_- \neq \bullet$ have positive conformal weights at the order $\mathcal{O} (c)$.

The coset models are equivalent to the minimal models $(p',  p)$.  For $N=2$,  the level $k$ is related with $(p',  p)$ through the relation
\be\label{eq:rel of k and p pm}
  k = \frac{3 p' - 2 p}{p - p'}\, ,
\ee
which for finite $p$ and large $p'$ becomes $k = - 3 - p / p' + \mathcal{O} \left((p')^{-2}\right)$.  Hence,  in the limit $p$ finite and $p' \to \infty$ we have $N=2$ and $k \to -3^{-}$.  This is exactly the limit for the semiclassical $W_2$ CFT.  More precisely,  the union of all the minimal models $(p',  p)$ with $p$ finite and $p' \to \infty$ is the semiclassical $W_2$ CFT.

For the enstrophy cascade of the 2d turbulence,  instead of the minimal models in a limit,  we find an equivalent description from the $W_2$ CFT at the level $k=-3^-$ in the semiclassical limit.  This CFT also provides exactly the Kraichnan-Batchelor scaling $k^{-3}$.  For the semiclassical $W_2$ CFT at $k = -3^-$ and $N = 2$, the primary operators are denoted by $(\Lambda_+,\, \Lambda_-)$, where $\Lambda_+$ and $\Lambda_-$ are representations of $\mathfrak{su} (2)_{-3}$ and $\mathfrak{su} (2)_{-2}$ respectively, which can be labeled by Young tableaux with boxes only in one row.  Using these notations,  we see that the enstrophy cascade of 2d turbulence can be realized with
\be
  \psi = \left(\,\begin{tiny}\ydiagram{2}\end{tiny}\, ,\,\, \bullet \right)\, ,\quad \phi = \left(\,\begin{tiny}\ydiagram{4}\end{tiny}\, ,\,\, \bullet \right)\, ,
\ee
because the $\Lambda_-$ is just the identity,  and the fusion rule of $\psi\times \psi$ can be obtained by only considering the tensor product of the $\Lambda_+$ part:
\begin{align}
  \begin{tiny}\ydiagram{2}\end{tiny}\otimes \begin{tiny}\ydiagram{2}\end{tiny} & = \bullet\,\, \oplus\,\, \begin{tiny}\ydiagram{2}\end{tiny}\,\, \oplus\,\, \begin{tiny}\ydiagram{4}\end{tiny} \,\, .
\end{align}
The operators $\psi$ and $\phi$ have the conformal weights
\be
  h_\psi = -1 - \frac{12}{c} + \mathcal{O} (c^{-2})\, ,\,\, h_\phi = -2 - \frac{36}{c} + \mathcal{O} (c^{-2})\, ,
\ee
and $\phi$ has the vacuum expectation value $\langle \phi \rangle = 0$,  which precisely leads to the scaling $E(k) \sim k^{-3 + \mathcal{O} (c^{-1})}$ and the constraint $h_\phi - 2 h_\psi = -12/c > 0$ for $c < 0$.   The primary operators $(r,  s)$ in the minimal model description and $(r_+,  r_-)$ in the semiclassical $W_2$ CFT are related through the identifications
\be
  r = r_+ + 1\, ,\quad s = r_- + 1\, .
\ee
With these relations,  the minimal models $(p',  p)$ in the limit $p$ finite,  $p' \to \infty$ and the semiclassical $W_2$ CFT lead to the same result.

\section*{Appendix B: Summary of 2D $O(N)$ Model and $Q$-Potts Model}\label{app:O(N) and Q-Potts}

This appendix summarizes some facts about the 2d $O(N)$ and $Q$-Potts models, which are relevant to the energy cascade of 2d turbulence. In particular, we will discuss some dualities between these two models and clarify some confusions in the literature.

\subsection*{2D $O(N)$ Model}
The $d$-dimensional $O(N)$ model can be defined as a Landau-Ginzburg type quantum field theory with $N$ commuting scalar fields as
\be\label{eq:O(N) model action}
  S = \int d^d x\, \left[\frac{1}{2} \sum_{a=1}^N \left(\partial \phi_a \right)^2 + \frac{1}{2} m_0^2 \sum_{a=1}^N \phi_a^2 + \lambda \sum_{a, b = 1}^N \phi_a^2\, \phi_b^2 \right]\, .
\ee
The primary operators of the $O(N)$ model can be expressed in terms of these scalar fields.  For instance,  the bulk energy operator in the $O(N)$ model is $\epsilon \equiv \sum_{a=1}^N : \phi_a^2:$,  which can be identified as the Kac operator $\phi_{1,  3}$ in minimal models.

The 2d $O(N)$ model can also be defined as a lattice model.  On the honeycomb lattice, it has the partition function
\be
  Z = \int \prod_{i=1}^N \prod_{\langle i, j \rangle} \left(1 + K\, S_i\cdot S_j \right)\, ,
\ee
where $S_i$ are $N$-vectors with $S_i^2 = N$,  and $\langle i,  j \rangle$ denotes the nearest neighbour sites.  One can evaluate this partition function as
\be
  Z = \sum_{\text{nonintersecting loops}} K^{n_B}\, N^{n_L}\, ,
\ee
where $n_B$ and $n_L$ stand for the total numbers of bonds and loops, respectively.  The high temperature corresponds to small $K$,  for which the $O(N)$ model becomes a geometrical loop model.  There is a critical point given by
\be
  K_c = \left(2 + \sqrt{2 - N} \right)^{- \frac{1}{2}}\, ,\quad \text{for } N \in [-2,\, 2]\, .
\ee
The phase at the critical point $K = K_c$ is called the dilute phase.  For any value of $K > K_c$, there is another phase called the dense phase,  which corresponds to infinite loops or lines filling the lattice with a finite nonzero density.

At $K \geq K_c$ the 2d $O(N)$ model becomes a CFT,  which can be described by the Coulomb gas.  The central charge of the 2d $O(N)$ model CFT is
\be\label{eq:CentralCharge}
  c = 1 - 6\, \frac{(g - 1)^2}{g}\, ,
\ee
where the parameter $g$ is the Coulomb gas coupling constant,  which is related to the parameter $N$ as follows:
\be
  N = - 2\, \textrm{cos} (\pi g)\, .
\ee
Different ranges of the parameter $g$ are chosen to correspond to different phases.   For the dilute phase ($K = K_c$) the parameter $g$ is chosen to be
\be
  g \in [1,\, 2]\, ,
\ee
while for the dense phase ($K > K_c$) the range of $g$ is
\be
  g \in [0,\, 1]\, .
\ee
As special cases,  $g = \frac{3}{2}$ in the dilute phase and $g = \frac{2}{3}$ in the dense phase both lead to the central charge $c = 0$,  and $g \to 0^+$ corresponds to the classical limit $c \to - \infty$.  In the Coulomb gas formalism,  the primary operators of the $O(N)$ model has both electric and magnetic charges $(e,  m)$.  The conformal weights of a primary operator with $(e,  m)$ are
\be
  h = x_{em} + \frac{c-1}{24}\, ,\quad \bar{h} = \bar{x}_{em} + \frac{c-1}{24}\, ,
\ee
where
\be
  x_{em} \equiv \frac{1}{4} \left(\frac{e}{\sqrt{g}} + m \sqrt{g} \right)^2\, ,\quad \bar{x}_{em} \equiv \frac{1}{4} \left(\frac{e}{\sqrt{g}} - m \sqrt{g} \right)^2\, .
\ee

The 2d $O(N)$ model can also be obtained from a minimal model $(p',\, p)$.  We know that the minimal models are unitary only when $|p' - p| = 1$.  Hence,  for unitary minimal models,  we choose $p' > p$ and adopt a new parametrization:
\be
  (p',  p) = (m+1,  m)\, ,\quad m \in \mathbb{Z}_+\, .
\ee
For the dilute phase with $g \in [1,  2]$,  the parameters have the relation
\be
  g = \frac{p'}{p} = \frac{m+1}{m} \quad \Rightarrow\quad m = \frac{1}{g - 1}\, .
\ee
For the dense phase with $g \in [0,  1]$,  the parameters have the relation
\be
  g = \frac{p}{p'} = \frac{m}{m+1} \quad\Rightarrow\quad m = \frac{g}{1 - g}\, .
\ee

The 2d $O(N)$ model can describe some physical systems.  For instance,  $N=2$ corresponds to the XY model at the Kosterlitz-Thouless transition point,  while the dilute phase of the $O(1)$ model is just the Ising model.  The 2d $O(N)$ model with $N = 0$ describes the self-avoiding random walk and the polymer at the critical point.  Some well-known realizations of the $O(N)$ model are summarized in Tab.~\ref{tab:O(N) models}.
\begin{table}[htb!]
\centering
\begin{tabular}{|c|c|c|}
\hline
$O(N)$ model & dilute phase ($1 \leq g \leq 2$) & dense phase ($0 \leq g \leq 1$)\\
{} & (for $1 \leq g \leq \frac{3}{2}$ dual to & (for $\frac{1}{2} \leq g \leq 1$ dual to\\
{} & $Q = N^2$ tricritical & $Q = N^2$ critical\\
{} & Potts model) & Potts model)\\
\hline \hline
$N=2$ &  \multicolumn{2}{|c|}{XY model at the Kosterlitz-Thouless}\\
{} & \multicolumn{2}{|c|}{transition point ($g = 1$,  $c = 1$)}
\\ \hline
$N=\sqrt{2}$ & $O(\sqrt{2})$ model & ($Q$=2)-Potts model log CFT\\
{} & ($g = \frac{5}{4}$,  $c = \frac{7}{10}$) & ($g = \frac{3}{4}$,  $c = \frac{1}{2}$)
\\ \hline
$N=1$ & Ising model & ($Q$=1)-Potts model log CFT\\
{} & ($g = \frac{4}{3}$,  $c = \frac{1}{2}$) & ($g = \frac{2}{3}$,  $c = 0$)
\\ \hline
$N=0$ & $O(0)$ model log CFT & ($Q$=0)-Potts model log CFT \\
{} & dilute polymer & dense polymer \\
{} & ($g = \frac{3}{2}$,  $c = 0$) & ($g = \frac{1}{2}$,  $c = -2$)
\\ \hline \hline
$N = -2$ & $O(-2)$ model log CFT & classical limit\\
{} & ($g = 2$,  $c = -2$) & ($g \to 0^+$,  $c \to - \infty$)
\\ \hline
\end{tabular} 
\caption{Dilute and dense phases of the $O(N)$ model at some special values of $N$\label{tab:O(N) models}}
\end{table}

For $-2 \leq N \leq 2$,  the 2d $O(N)$ model on a torus has the partition function \cite{1987diFrancesco}:
\be\label{eq:O(N) Z}
  Z_N (g,\, e_0; \tau) = \sum_{m,  m' \in \mathbb{Z}} \textrm{cos} \Big[\pi\, e_0\, \textrm{gcd} (|m|,\, |m'|) \Big]\, Z_{m,  m'} \left(\frac{g}{4}; \tau\right)\, ,
\ee
with $\textrm{gcd} (|m|,\, |m'|)$ denoting the greatest common divisor of two integers $|m|$ and $|m'|$,  and
\be\label{eq:O(N) Zmm'}
  Z_{m,  m'} (g; \tau) = \frac{g}{|\eta(\tau)|^2\, \sqrt{\textrm{Im}\, \tau}}\, \textrm{exp} \left[- \frac{\pi g^2 |m \tau - m'|^2}{\textrm{Im}\, \tau} \right]\, .
\ee
Suppose that the torus is defined by a rectangle of size $L \times T$ with periodic boundary conditions.  The parameter $\tau \equiv i T / L$ with $L, \, T \to \infty$.  Moreover,  the parameter $e_0$ is defined by
\be
  2\, \textrm{cos} (\pi\, e_0) = N\, ,
\ee
with $0 \leq e_0 \leq 1$.  Evaluating the $O(N)$ model's partition function for $N=0$,  we obtain the partition function for the $O$($N$=0) model:
\be
  Z_{N=0} = 1\, ,
\ee
which looks trivial.  However,  the derivative of the partition function at $N=0$ is still nontrivial.

\subsection*{2D $Q$-Potts Model}

The Potts model is a generalization of the Ising model,  in which the spin can take $Q$ possible states.  The Hamiltonian of the $Q$-Potts model is
\be
  H = - J\, \sum_{\langle i,  j \rangle} \delta_{s (i),  s(j)}\, .
\ee
Consequently,  the partition function of the $Q$-Potts model is
\be
  Z (Q,  x) = \textrm{Tr}_s \prod_{\langle i,  j \rangle}  \left(1 + x\, \delta_{s (i),  s(j)} \right)\, ,
\ee
where the parameter $x$ is defined as
\be
  x \equiv e^J - 1\, .
\ee
When defining on a lattice with $N$ nearest neighbour bonds,  the partition function of the $Q$-Potts model has another expression:
\be
  Z (Q,  x) = (1 + x)^{-N}\, \sum_{\mathcal{C}} Q^{n_c}\, x^{n_b}\, ,
\ee
where the summation is taken over distinct cluster configurations,  while $n_c$ and $n_b$ stand for the total number of clusters and the total number of bonds in the configuration respectively.

The $Q$-Potts model also has a Landau-Ginzburg type field theory description.  Let us introduce a continuous spin variable $S_a (r)$ with the constraint $\sum_{a=1}^Q S_a = Q - 1$.  Furthermore,  we define an order parameter
\be\label{eq:Potts constraint}
  \phi_a \equiv S_a - \frac{1}{Q} \sum_a S_a\, ,\quad \text{with}\quad \sum_{a=1}^Q \phi_a = 0\, .
\ee
The $d$-dimensional $Q$-Potts model is then given by the action
\be
  S = \int d^d x\, \left[\sum_a \left(\partial \phi_a \right)^2 + m_0^2 \sum_a \phi_a^2 + \lambda \sum_a \phi_a^3 + \mathcal{O} (\phi^4) \right]\, ,
\ee
with the constraint given above.

The Coulomb gas formalism can also be used to study the 2d $Q$-Potts model.  The central charge $c$ is given by the Coulomb gas coupling constant $g$ through
\be
  c = 1 - \frac{6\, (2 - g/2)^2}{g}\, .
\ee
The parameter $Q$ and the coupling constant $g$ can be related by
\be
  Q = 2 + 2\, \textrm{cos} \left(2 \pi g \right) = 4\, \textrm{cos}^2 \left(\pi g / 2 \right)\, ,\quad g \in \left[2,\, 4 \right]\, .
\ee
The parameter $e_0$ is related to $Q$ via
\be
  \sqrt{Q} = 2\, \textrm{cos} (\pi e_0 / 2)\, .
\ee
Hence,  for the $Q$-Potts model
\be
  e_0 = \pm (2 - g / 2)\,\, \textrm{mod}\,\, 4\, .
\ee

For $0 \leq Q \leq 4$,  the 2d $Q$-Potts model's torus partition function is related to the 2d $O(N)$ model's torus partition function:
\be\label{eq:Q-Potts Z}
  Z_Q = Z_N (g,\, e_0) + \frac{1}{2} (Q-1)\, \Big(Z_c [g,\, 1] - Z_c [g,\, 1/2] \Big)\, ,
\ee
where $Z_c [g,\, 1]$ is the modular invariant Coulombic partition function defined by
\be
  Z_c [g,  1] \equiv \sum_{m,  m' \in \mathbb{Z}}  Z_{m,  m'} (g)\, ,
\ee
which has the symmetries
\be
  Z_c [g,  f] = Z_c [1/g,  1/f] = Z_c [g f^2,  1]\, .
\ee
Evaluating the $Q$-Potts model's partition function for $Q=1$,  we obtain the partition function for the ($Q$=1)-Potts model
\be
  Z_{Q=1} = 1\, ,
\ee
which looks trivial.  Again,  the derivative of this partition function at $Q=1$ is nontrivial.

\subsection*{Relations Between 2D $O(N)$ and $Q$-Potts Models}

There are some dualities between the 2d $O(N)$ model and the 2d $Q$-Potts model.  In this section, let us make a summary of these dualities.

As we have already mentioned in Tab.~\ref{tab:O(N) models},  the dilute phase of the $O(N)$ model is dual to the tricritial $Q$-Potts model,  while the dense phase of the $O(N)$ model is dual to the critical $Q$-Potts model.  More precisely,  in the Coulomb gas formalism for $\frac{1}{2} \leq g \leq 1$:
\be
  \text{2d $O(N)$ model dense phase} \,\,\xLeftrightarrow{N = \sqrt{Q}}\,\, \text{2d critical $Q$-Potts model}\, ,
\ee
while for $1 \leq g \leq \frac{3}{2}$:
\be
  \text{2d $O(N)$ model dilute phase} \,\,\xLeftrightarrow{N = \sqrt{Q}}\,\, \text{2d tricritical $Q$-Potts model}\, .
\ee
For a domain with a boundary,  these two dualities hold both on the boundary and in the bulk.

The $c=0$ models are a little special,  and this class includes several different models.  The 2d $O$($N$=0) model and the 2d ($Q$=1)-Potts model both belong to the $c=0$ models,  and both of them are logarithmic CFTs.  In order to distinguish different $c=0$ CFTs,  one can define a new parameter $b$.  Assume that there exists an operator $\widetilde{T}$ with the conformal weights $(2 + \bar{h}, \, \bar{h})$,  which collides with the energy-momentum tensor $T$ in the limit $c \to 0$,  i.e., $\bar{h} \to 0$ when $c \to 0$.  The parameter $b$ is defined as
\be
  b \equiv - \frac{1}{2} \lim_{c \to 0} \frac{c}{\bar{h}} \quad\text{or}\quad b \equiv - \frac{1}{\left(dx_{\widetilde{T}} /  dc \right)_{c = 0}}\, ,
\ee
where $x_{\widetilde{T}}$ is the scaling dimension of $\widetilde{T}$.  This parameter was first defined in \cite{Gurarie:1999bp}.  Later in \cite{Gurarie:1999yx},  it was conjectured that $b$ takes the value $- \frac{5}{8}$ for the 2d $O$($N$=0) model,  and the value $\frac{5}{6}$ for the 2d ($Q$=1)-Potts model.  However,  more studies show that these preliminary results were incorrect.  The correct values of $b$ also differ for boundary and bulk.

On the boundary,  the $\widetilde{T}$ operator is $\phi_{3,1}$ for the 2d $O$($N$=0) model,  which has the conformal weight $2 - 3 c / 5$,  and $\phi_{1,5}$ for the 2d ($Q$=1)-Potts model,  which has the conformal weight $2 + 4 c / 5$.  Hence,  the boundary $b$ parameter takes the values \cite{Dubail:2010zz,  Vasseur:2011fi,  Mathieu:2007pe}
\be
  b_{\text{bdy}} = \left\{
  \begin{aligned}
    & \frac{5}{6}\, , & \text{for the 2d $O$($N$=0) model} \, ; \\
    & - \frac{5}{8}\, , & \text{for the 2d ($Q$=1)-Potts model}\, .
  \end{aligned} \right.
\ee
However,  the Kac operator $\phi_{2,1}$ is the boundary condition changing operator \cite{QuantumImpurityProblem} in the ($Q$=1)-Potts model.  We expect that when $\phi_{2,1}$ acting on a boundary Cardy state should lead to
\be\label{eq:byd cond changing op for Potts}
  \phi_{2,1} |\, \widetilde{\phi}_{1,5}\, \rangle = |\, \widetilde{\phi}_{3,1}\, \rangle\, .
\ee
Consequently,  after inserting $\phi_{2,1}$ on the boundary,  the $\widetilde{T}$ for the ($Q$=1)-Potts model is $\phi_{3,1}$ instead of $\phi_{1,5}$,  and the corresponding $b$-parameter for the ($Q$=1)-Potts model in the presence of the boundary condition changing operator $\phi_{2,1}$ is $\frac{5}{6}$ instead of its original value $-\frac{5}{8}$.  Therefore,  the ($Q$=1)-Potts model with a boundary condition changing operator $\phi_{2,1}$ and the $O$($N$=0) model have the same boundary $b$-parameter.

In the bulk,  the $\widetilde{T}$ operator is a spin-2 2-leg operator for the $O$($N$=0) model,  which has the scaling dimension $1 + 3 / (2 g)$,  and a spin-2 4-leg (or equivalently 2-cluster) operator for the ($Q$=1)-Potts model,  which has the scaling dimension $1 + 3 g / 2$.  Hence,  the bulk parameter $b$ takes the values \cite{Vasseur:2011ud}
\be
  b_{\text{bulk}} = \left\{
  \begin{aligned}
    & -5\, , & \text{for the 2d $O$($N$=0) model} \, ; \\
    & -5\, , & \text{for the 2d ($Q$=1)-Potts model}\, .
  \end{aligned} \right.
\ee
Because the 2d $O$($N$=0) model and the 2d ($Q$=1)-Potts model can have the same values of the central charge $c$ and the $b$ parameter both in the bulk and on the boundary,  they can coexist and form a direct sum CFT:
\be
  \Big(\text{2d $O$($N$=0) model}\Big) \oplus \Big(\text{2d ($Q$=1)-Potts model}\Big)\, ,
\ee
which is relevant to the energy cascade of 2d turbulence.

These two $c=0$ logarithmic CFTs in the bulk of a domain are related by the SLE relation \cite{Duplantier:1999cz}:
\be
  \text{2d $O$($N$=0) model} \quad\xLeftrightarrow{\text{SLE relation}}\quad \text{2d ($Q$=1)-Potts model}\, .
\ee
The name SLE relation originates from that both of these 2d CFTs are related to the Schramm-Loewner evolution (SLE).   A 2d SLE is given by a Brownian walk $\xi (t)$ satisfying
\be
  \langle \left(\xi (t) - \xi (0) \right)^2 \rangle = \kappa t\, .
\ee
For an SLE$_\kappa$ curve,  the fractal dimension is
\be
  D_\kappa = 1 + \frac{\kappa}{8}\quad \text{for } \kappa < 8\, .
\ee
A 2d SLE also implies an underlying CFT with the central charge
\be
  c = \frac{(8 - 3 \kappa) (\kappa - 6)}{2 \kappa}\, .
\ee
The SLE relation connects SLE$_\kappa$ and SLE$_{16 / \kappa}$ ($\kappa > 4$).  A special case is
\be
  \text{SLE}_6 \quad \xLeftrightarrow{\text{SLE relation}} \quad \text{SLE}_{\frac{8}{3}}\, ,
\ee
which correspond to the 2d ($Q$=1)-Potts model and the 2d $O$($N$=0) model respectively. However,  the SLE relation does not hold on the boundary,  which is consistent with the nonunique boundary values of the $b$ parameter.

\section*{Appendix C: OPE of the Boundary Logarithmic $O$($N$=0) Model}\label{app:OPE log O(N)}

In the literature,  the fusion rule of a CFT is usually presented as a decomposition in short-handed notation.  For instance,  in a minimal model $(p',\, p)$ the fusion of two primary Kac operators is
\be\label{eq:fusion rule}
  \phi_{r, s} \times \phi_{r',  s'} = \sum_{(r'',\, s'')} \phi_{r'', s''}\, .
\ee
This fusion rule is a purely algebraic relation, including only primary operators without descendants.  In contrast to the fusion rule,  the operator product expansion (OPE) also deals with descendants.  For instance,  if we only focus on the identity operator in the fusion channel,  there can still be the descendants of the identity in the OPE \cite{Cardy:2013rqg}:
\be
  \Phi(z,  \bar{z})\cdot \Phi(0,0) = \frac{a}{z^{2 h_\phi}\, \bar{z}^{2 \bar{h}_\phi}} \Bigg[I + \frac{2 h_\phi}{c} z^2 T(0) + \frac{2 \bar{h}_\phi}{c} \bar{z}^2 \overline{T} (0) + \frac{4 h_\phi \bar{h}_\phi}{c^2} z^2 \bar{z}^2 T \overline{T} (0) + \cdots \Bigg]\, .
\ee
In general,  we should consider both kinds of expansions shown above.

In this appendix,  we focus on the following fusion rule of primary operators in the bulk $O(N)$ model:
\be
  \phi_{1,3} \times \phi_{1, 3} = \phi_{1,1} + \phi_{1,3} + \phi_{1,5} \equiv I + \psi + \phi_{1,5}\, .
\ee
Taking into account the contributions from the descendants,  we obtain in the bulk:
\begin{align}
  {} & \psi (z,  \bar{z}) \cdot \psi (0, 0) \nonumber\\
  = & \, \phi_{1,3} (z,  \bar{z}) \cdot \phi_{1,3} (0, 0) \nonumber\\
  = & \, \frac{a_I}{z^{2 h_\psi}\, \bar{z}^{2 \bar{h}_\psi}} \Bigg[I + \frac{2 h_\psi}{c} z^2 T(0,  0) + \frac{2 \bar{h}_\psi}{c} \bar{z}^2 \overline{T} (0,  0) + \frac{4 h_\psi \bar{h}_\psi}{c^2} z^2 \bar{z}^2 T \overline{T} (0,  0) + \cdots \Bigg] \nonumber\\
  {} & \, + \frac{a_\psi }{z^{h_\psi}\, \bar{z}^{\bar{h}_\psi}} \Big[\psi (0, 0) + \cdots \Big] + \frac{a_\phi}{z^{2 h_\psi - h_\phi}\, \bar{z}^{2 \bar{h}_\psi - \bar{h}_\phi}} \Big[\phi_{1,5} (0, 0) + \cdots \Big]\, ,\label{eq:full OPE}
\end{align}
where the dots denote the higher-order descendants.

For the $O$($N$=0) model,  two operators in the OPE above,  $T \overline{T}$ and $\phi_{1,5}$,  have the same conformal weights $(2,  2)$.  Moreover,  it is shown in \cite{Cardy:2013rqg} that the coefficient $a_I \propto N$,  hence $a_I$ vanishes when $N=0$.  We can analyze the central charge $c$ in the small-$N$ limit and find that they are related in the following way:
\be
  c = \frac{5 N}{3 \pi} + \mathcal{O} (N^2)\, .
\ee
At the order shown in the OPE above,  there is a problem caused by the term $\sim T \overline{T} (0)$,  which is proportional to $a_I / c^2$ and diverges in the limit $N \to 0$.  This is the so-called $c \to 0$ paradox.  To resolve this paradox,  we first assume that $a_{\psi, \, \phi} \sim \mathcal{O} (c^0)$,  which can be realized by redefining the primary operators $\psi$ and $\phi_{1,5}$.  Next,  we focus on the two colliding terms,  i.e.,  the two terms with the same conformal weights.  Suppose that $c$ is a little away from $0$,  and then the conformal weights of $\phi$ are $(2 + \delta,  2 + \delta)$,  where $\delta \sim \mathcal{O} (c)$.  More precisely,  we find for $\phi \equiv \phi_{1,5}$ that
\be
  \delta =\frac{4}{5}\, c\, .
\ee
Consequently,  the two colliding terms in the OPE become
\begin{align}
  {} & \,\, \frac{a_I}{z^{2 h_\psi}\, \bar{z}^{2 \bar{h}_\psi}} \frac{4 h_\psi \bar{h}_\psi}{c^2} z^2 \bar{z}^2 T \overline{T} (0,  0) + \frac{a_\phi}{z^{2 h_\psi - h_\phi}\, \bar{z}^{2 \bar{h}_\psi - \bar{h}_\phi}} \phi_{1,5} (0, 0) \nonumber\\
  = & \,\, \frac{1}{z^{2 h_\psi}\, \bar{z}^{2 \bar{h}_\psi}} \Bigg[ a_I \frac{4 h_\psi \bar{h}_\psi}{c^2} z^2 \bar{z}^2 T \overline{T} (0,  0) + a_\phi\, z^{2 + \delta} \bar{z}^{2 + \delta} \phi_{1,5} (0, 0) \Bigg] \nonumber\\
  = & \,\, \frac{1}{z^{2 h_\psi}\, \bar{z}^{2 \bar{h}_\psi}} \Bigg[ \bigg(a_I 4 h_\psi \bar{h}_\psi T \overline{T} (0,  0) + c^2 a_\phi\, \phi_{1,5} (0, 0) \bigg) \frac{z^{2 + \delta} \bar{z}^{2 + \delta}}{c^2} + a_I \frac{4 h_\psi \bar{h}_\psi}{c^2} T \overline{T} (0,  0) \bigg(z^2 \bar{z}^2 - z^{2 + \delta} \bar{z}^{2 + \delta} \bigg) \Bigg] \nonumber\\
  = & \,\, \frac{1}{z^{2 h_\psi}\, \bar{z}^{2 \bar{h}_\psi}} \Bigg[ \frac{a_I}{\delta} h_\psi \bar{h}_\psi \frac{T \overline{T} (0,  0) + \frac{c^2 a_\phi\, \phi_{1,5} (0,  0)}{a_I 4 h_\psi \bar{h}_\psi}}{\delta}\, \frac{4 \delta^2}{c^2}\, z^{2 + \delta} \bar{z}^{2 + \delta} + a_I h_\psi \bar{h}_\psi \frac{4 \delta^2}{c^2}\, \frac{z^2 \bar{z}^2 - z^{2 + \delta} \bar{z}^{2 + \delta}}{\delta^2}\, T \overline{T} (0,  0) \Bigg]\, .\label{eq:colliding terms 1}
\end{align}
In the limit $\delta \to 0$ or equivalently $c \to 0$,  we define two new parameters
\be
  \widetilde{a}_I \equiv \lim_{\delta \to 0} \frac{a_I}{\delta}\, ,\quad b \equiv \lim_{\delta \to 0} \frac{c}{2 \delta}\, .
\ee
We also define a new operator
\be
  \phi \equiv \frac{T \overline{T} + \frac{c^2 a_\phi\, \phi_{1,5}}{a_I 4 h_\psi \bar{h}_\psi}}{\delta} = \frac{\delta}{a_I}\, \frac{c^2}{4 \delta^2}\, \frac{a_\phi}{4 h_\psi \bar{h}_\psi}\, \phi_{1,5} + \frac{1}{\delta} T \overline{T}\, ,
\ee
which in the limit $\delta \to 0$ or equivalently $c \to 0$ becomes
\be\label{eq:define phi}
  \phi = C_\phi\cdot \phi_{1,5} + \frac{1}{\delta} T \overline{T}\quad \text{with}\quad C_\phi \equiv \frac{b^2}{\widetilde{a}_I}\, \frac{a_\phi}{4 h_\psi \bar{h}_\psi}\, .
\ee
Using these definitions,  we can rewrite the two colliding terms in the limit $\delta \to 0$ as
\begin{align}
  {} & \,\, \frac{a_I}{z^{2 h_\psi}\, \bar{z}^{2 \bar{h}_\psi}} \frac{4 h_\psi \bar{h}_\psi}{c^2} z^2 \bar{z}^2 T \overline{T} (0,  0) + \frac{a_\phi}{z^{2 h_\psi - h_\phi}\, \bar{z}^{2 \bar{h}_\psi - \bar{h}_\phi}} \phi_{1,5} (0, 0) \nonumber\\
  = & \,\, \frac{1}{z^{2 h_\psi}\, \bar{z}^{2 \bar{h}_\psi}} \Bigg[ \widetilde{a}_I\, h_\psi \bar{h}_\psi\, \frac{1}{b^2}\, z^2 \bar{z}^2\, \phi (0,  0) - a_I\, h_\psi \bar{h}_\psi\, \frac{1}{b^2}\, z^2 \bar{z}^2\, \textrm{log} (z \bar{z})\, T \overline{T} (0,  0) \Bigg]\, ,\label{eq:colliding terms 2}
\end{align}
where the first term is the leading contribution of the order $\mathcal{O} (c^0)$,  while the second term of the order $\mathcal{O} (c)$ is subleading,  because
\be
  \widetilde{a}_I \sim \mathcal{O} (c^0)\, ,\quad b \sim \mathcal{O} (c^0)\, ,\quad a_I \sim \mathcal{O} (c)\, .
\ee
Hence,  after taking into account the logarithmic pair $(\phi_{1,5},\, T \overline{T})$,  the full OPE in the limit $c \to 0$ becomes
\begin{align}
  {} & \psi (z,  \bar{z}) \cdot \psi (0, 0) \nonumber\\
  = & \, \frac{\widetilde{a}_I}{z^{2 h_\psi}\, \bar{z}^{2 \bar{h}_\psi}} \Bigg[\frac{h_\psi}{b} z^2 T(0,  0) + \frac{\bar{h}_\psi}{b} \bar{z}^2 \overline{T} (0,  0) + \frac{h_\psi \bar{h}_\psi}{b^2} z^2 \bar{z}^2\, \phi (0,  0) + \cdots \Bigg] \nonumber\\
  {} & \, + \frac{a_\psi }{z^{2 h_\psi}\, \bar{z}^{2 \bar{h}_\psi}} \bigg[z^{h_\psi}\, \bar{z}^{\bar{h}_\psi}\, \psi (0, 0) + \cdots \bigg] + \mathcal{O} (c)\, ,\label{eq:full OPE after log pair}
\end{align}
and the $c \to 0$ paradox is resolved at this order.  A term $\sim z^2 \bar{z}^2\, \textrm{log} (z \bar{z})$ emerges at the order $\mathcal{O} (c)$.  At higher orders,  more divergent terms will appear,  and the $c \to 0$ paradox can be resolved in a similar manner.

This paper also needs to consider the OPE of three operators $\psi \cdot \psi \cdot \psi$.  To simplify the discussion,  we first use the fusion rule to two of the three operators at the same point,  then apply the full OPE to the fusion products with the third operator at another point.  Since the boundary operators constrain the bulk operator content,  when two $\phi_{1,3}$ are inserted at two ends of a cylinder,  the allowed operators in the OPE $\psi \cdot \psi \cdot \psi$ are still $\phi_{1,1}$,  $\phi_{1,3}$ and $\phi_{1,5}$.  Hence,  the 3-operator OPE still maintains the same form as the 2-operator OPE but with different coefficients:
\begin{align}
  {} & \psi (z,  \bar{z}) \cdot \psi (0, 0) \cdot \psi (0, 0) \nonumber\\
  = & \, \phi_{1,3} (z,  \bar{z}) \cdot \phi_{1,3} (0, 0) \cdot \phi_{1,3} (0, 0) \nonumber\\
  = & \, \frac{a'_I}{z^{2 h_\psi}\, \bar{z}^{2 \bar{h}_\psi}} \Bigg[I + \frac{2 h_\psi}{c} z^2 T(0,  0) + \frac{2 \bar{h}_\psi}{c} \bar{z}^2 \overline{T} (0,  0) + \frac{4 h_\psi \bar{h}_\psi}{c^2} z^2 \bar{z}^2 T \overline{T} (0,  0) + \cdots \Bigg] \nonumber\\
  {} & \, + \frac{a'_\psi }{z^{h_\psi}\, \bar{z}^{\bar{h}_\psi}} \Big[\psi (0, 0) + \cdots \Big] + \frac{a'_\phi}{z^{2 h_\psi - h_\phi}\, \bar{z}^{2 \bar{h}_\psi - \bar{h}_\phi}} \Big[\phi_{1,5} (0, 0) + \cdots \Big]\, .
\end{align}
Applying the same procedure,  we find that the $c \to 0$ paradox in this 3-operator OPE can be resolved by defining a new operator
\be
  \chi \equiv \frac{\delta}{a'_I}\, \frac{c^2}{4 \delta^2}\, \frac{a'_\phi}{4 h_\psi \bar{h}_\psi}\, \phi_{1,5} + \frac{1}{\delta} T \overline{T}\, ,
\ee
which in the limit $\delta \to 0$ or $c \to 0$ approaches
\be
  \chi = C'_\phi\cdot \phi_{1,5} + \frac{1}{\delta} T \overline{T} \quad\text{with}\quad C'_\phi \equiv \frac{b^2}{\widetilde{a}'_I}\, \frac{a'_\phi}{4 h_\psi \bar{h}_\psi}\, ,
\ee
where $\widetilde{a}'_I \equiv \lim_{\delta \to 0} a'_I / \delta$,  and $C'_\phi$ is a constant generally different from $C_\phi$.

\section*{Appendix D: Scaling Dimensions of Boundary Operators}\label{app:BoundaryScalingDimensions}

For a boundary CFT,  the scaling dimensions of boundary operators are defined in the boundary directions.  The values of these boundary scaling dimensions can be computed using the method of images and are usually different from their bulk counterparts.  For instance,  in the 2d Ising CFT,  the boundary spin operator has the scaling dimension $\Delta_\sigma^{\text{boundary}} = \frac{1}{2}$ or $2$ depending on the boundary conditions \cite{Cardy:1984bb},  which is not the same as its bulk scaling dimension $\Delta_\sigma^{\text{bulk}} = \frac{1}{8}$.

Besides the scaling dimensions of boundary operators along the boundary direction,  sometimes we also need their effective scaling dimensions in the direction perpendicular to the boundary,  i.e.,  from the boundary towards the bulk.  The boundary operators are defined only in the boundary directions and have no dependence on the direction towards the bulk.  However,  we can use the bulk-boundary OPE discussed in \cite{Cardy:2004hm} first to express the boundary operators in terms of bulk operators and then discuss their effective scaling dimensions in the bulk.  The goal of this appendix is to find the scale dependence of the boundary correlation functions along the bulk direction,  e.g.,  $\langle \widetilde{\psi} (L)\, |\, \widetilde{\psi} (0) \rangle$.

The bulk operator in a boundary CFT can be treated using the method of images.  When the bulk operator approaches the boundary,  so does its image,  and they will eventually collide at the boundary and form an operator product expansion.  This OPE is equivalent to an OPE between bulk and boundary operators. Hence,  it is called the bulk-boundary OPE.  The bulk-boundary OPE has the general expression:
\be
  \phi_j (z_j,\, \bar{z}_j) = \sum_k d_{jk}\, y_j^{- h_j - \bar{h}_j + h_k}\, \widetilde{\phi}_k (x_j)\, ,
\ee
where $z_j \equiv x_j + i\, y_j$.  The operators with and without tilde denote the boundary and the bulk operators respectively.  The OPE coefficients $d_{jk}$ are nonvanishing only when the fusion coefficients $N_{h_j,\, \bar{h}_j}^{h_k} > 0$.  Conversely,  the boundary operator can also be written as an expansion of bulk operators:
\be
  \widetilde{\phi}_k (x_k) = \sum_j \widetilde{d}_{kj}\, y_k^{- h_k + h_j + \bar{h}_j}\, \phi_j (z_k,\, \bar{z}_k)\, .
\ee
For instance,  for the CFT description of the energy cascade of 2d turbulence we have inserted two operators $\widetilde{\psi} = \widetilde{\phi}_{1,3}$ at the boundary.  For this case,  we have
\begin{align}
  \widetilde{\psi} (x_k) & = \widetilde{d}_{\psi\psi}\, y_k^{h_{1,3}}\, \phi_{1,3} (z_k,\, \bar{z}_k) + \widetilde{d}_{\psi\phi}\, y_k^{2\, h_{1,5} - h_{1,3}}\, \phi_{1,5} (z_k,\, \bar{z}_k) \nonumber\\
  {} & = \widetilde{d}_{\psi\psi}\, y_k^{h_\psi}\, \psi (z_k,\, \bar{z}_k) + \widetilde{d}_{\psi\phi}\, y_k^{2\, h_\phi - h_\psi}\, \phi (z_k,\, \bar{z}_k)\, .
\end{align}
The bulk operator $\phi_{1,1}$ does not appear on the right-hand side of the expansion above,  because the matrix $\widetilde{d}_{kj}$ as the inverse of $d_{jk}$ has vanishing entries $\widetilde{d}_{\psi I} = 0$.  When the bulk operators are close to the boundary,  i.e., $y_k \to 0$,  the first term on the right-hand side in the expansion above dominates,  i.e.,
\be
  \widetilde{\psi} (x_k) \approx \widetilde{d}_{\psi\psi}\, y_k^{h_\psi}\, \psi (z_k,\, \bar{z}_k)\, .
\ee
Therefore,  for small $y_k$,
\begin{align}
  \langle \widetilde{\psi} (L)\, |\, \widetilde{\psi} (0) \rangle & \approx (\widetilde{d}_{\psi\psi})^2 (y_1\, y_2)^{h_\psi}\, \langle \psi (L)\, |\, \psi (0) \rangle = (\widetilde{d}_{\psi\psi})^2 \left(\frac{y_1\, y_2}{L^2} \right)^{h_\psi}\, L^{2\, h_\psi}\, \langle \psi (L)\, |\, \psi (0) \rangle \nonumber\\
  {} & \propto L^{2\, h_\psi}\, L^{- 4\, h_\psi} = L^{-2\, h_\psi}\, ,
\end{align}
where we dropped a dimensionless constant $\left(y_1\, y_2 / L^2 \right)^{h_\psi}$ in the last line.  This expression implies that the boundary operator $\widetilde{\psi}$ has an effective scaling dimension in the direction perpendicular to the boundary towards the bulk,  which is half of its bulk counterpart.

We can apply a similar analysis to higher-point correlation functions with boundary operators.  For instance,  the 1-point function with two boundary operators behaves like a 3-point function with particular conformal dimensions:
\be
  \langle \widetilde{\psi} (L)\, |\, \chi(0) \,|\, \widetilde{\psi} (0) \rangle \sim L^{- 2\, h_\psi}\, ,
\ee
where we used the general formula for the 3-point function of bulk operators
\be
  \langle \phi_1 (r_1)\, \phi_2 (r_2)\, \phi_3 (r_3) \rangle = \frac{C_{123}}{r_{12}^{\Delta_1 + \Delta_2 - \Delta_3}\, r_{13}^{\Delta_1 + \Delta_3 - \Delta_2}\, r_{23}^{\Delta_2 + \Delta_3 - \Delta_1}}\, .
\ee

The discussions above are valid for ordinary CFTs.  However, there is an additional subtlety for a $c=0$ logarithmic CFT.  As shown in App.~C,  the OPE may diverge for a $c=0$ logarithmic CFT, called the $c \to 0$ paradox.  This problem can be resolved by considering logarithmic pairs and working with some new operators.  Similar paradox and resolution also appear in the correlation functions.  For instance,  for the energy cascade discussed in the main text,  some intermediate steps of computing the energy flux without an external stirring force include
\begin{align}
  {} & \, \langle \widetilde{\psi} (L)\, |\, ({\bf u} \cdot \nabla {\bf u}) ({\bf k})\, {\bf u} (- {\bf k})\, |\, \widetilde{\psi} (0) \rangle \nonumber\\
  = & \, (k^{-1})^{- 2 h_{{\bf u} \cdot \nabla {\bf u}} - 2 h_{\bf u}}\, (k^{-1})^{2 h_\chi + 4}\, k^{-2}\, \langle \widetilde{\psi} (L)\, |\, (L_{-1} \overline{L}_{-2} L_{-1}\, \chi (0) + \cdots)\, |\, \widetilde{\psi} (0)\rangle \nonumber\\
  \sim & \, (k^{-1})^{- 2 h_{{\bf u} \cdot \nabla {\bf u}} - 2 h_{\bf u}}\, (k^{-1})^{2 h_\chi + 4}\, k^{-2}\, L^{- 2\, h_\psi}\, ,
\end{align}
where the factor $k^{-2}$ comes from the Fourier transforming the spatial 2-point function into the momentum space.  The operator $\chi$ has been defined in App.~C,  which is a linear combination of logarithmic pair $(\phi_{1,5},\,  T \overline{T})$.  Since the allowed bulk operators are restricted by the operators inserted on the boundary,  the multi-operator OPE is essentially the same as the OPE $\psi \cdot \psi$ in App.~C,  but with different coefficients.  Therefore,  the divergences in the multi-operator OPE and the corresponding multi-point correlation function can be removed by redefining some new operators from the logarithmic pair.  The OPE and correlation functions may have logarithmic terms,  but they show up at the subleading order $\mathcal{O} (c)$.

We can apply this result to compute the energy flux $J^{(E)}$ in the energy cascade at the IR cutoff $L^{-1}$ before introducing the external stirring force:
\begin{align}
  J^{(E)} (L^{-1}) = & \, - L^{2 (h_{\widetilde{\psi}} + h_{\widetilde{\psi}})} \int_{|{\bf k}| > L^{-1}} d^2 k\, \langle \dot{\bf u} ({\bf k})\, {\bf u} (- {\bf k}) \rangle_{\widetilde{\psi}\, \widetilde{\psi}} \nonumber\\
  \approx & \, - L^{2 (h_{\widetilde{\psi}} + h_{\widetilde{\psi}})} \int_{|{\bf k}| > L^{-1}} d^2 k\, \langle ({\bf u} \cdot \nabla {\bf u})\, {\bf u} \rangle_{\widetilde{\psi}\, \widetilde{\psi}} \nonumber\\
  = & \, - L^{2 (h_{\widetilde{\psi}} + h_{\widetilde{\psi}})} \int_{|{\bf k}| > L^{-1}} d^2 k\, \Big\langle \left( \epsilon_{\alpha\beta} (\partial_\beta \psi)\, \partial_\alpha (\epsilon_{\gamma\delta} \partial_\delta \psi) \right)\, \epsilon_{\gamma\rho} \partial_\rho \psi \Big\rangle_{\widetilde{\psi}\, \widetilde{\psi}} \nonumber\\
  \sim & \, - L^{4 h_{\widetilde{\psi}}}\, L^{- 2 h_{\bf u} - 2 h_{{\bf u} \cdot \nabla {\bf u}}}\, L^{2 h_\chi + 4}\, L^{- 2 h_{\widetilde{\psi}}} = - L^{2 h_{\widetilde{\psi}}}\, L^{- 2 h_{\bf u} - 2 h_{{\bf u} \cdot \nabla {\bf u}}}\, L^{2 h_\chi + 4}\nonumber\\
  = & \, - L^{2 h_\psi}\, L^{- (2 h_\psi + 1) - (2 h_\phi + 3)}\, L^{2 h_\chi + 4} = - L^{- 2 h_\phi + 2 h_\chi}\, ,
\end{align}
where we neglected a boundary term contribution from $- (\nabla p) / \rho$ in the NSE of velocity due to the fixed boundary condition ${\bf u}_\sigma |_{\sigma = 0,\, L} = 0$.  The $L$-independence of $J^{(E)}$ requires that $h_\chi = h_\phi$,  which helps determine the operator $\chi$ in the fusion products of $\psi \times \psi \times \psi$.  We can also compute the enstrophy flux $J^{(H)} (L^{-1})$ for the energy cascade in the absence of a stirring force,  and the scaling of $J^{(H)} (L^{-1})$ is the same as $J^{(E)} (L^{-1})$ shown above.

We emphasize that the effective scaling dimension for a boundary operator discussed in this appendix is generally not the same as the scaling dimension in the boundary direction.  This fact shows that a boundary operator may behave differently in different directions.

% For your review copy (i.e., the file you initially send in for
% evaluation), you can use the {figure} environment and the
% \includegraphics command to stream your figures into the text, placing
% all figures at the end.  For the final, revised manuscript for
% acceptance and production, however, PostScript or other graphics
% should not be streamed into your compliled file.  Instead, set
% captions as simple paragraphs (with a \noindent tag), setting them
% off from the rest of the text with a \clearpage as shown  below, and
% submit figures as separate files according to the Art Department's
% instructions.

% \clearpage

% \noindent {\bf Fig. 1.} Please do not use figure environments to set
% up your figures in the final (post-peer-review) draft, do not include graphics in your
% source code, and do not cite figures in the text using \LaTeX\
% \verb+\ref+ commands.  Instead, simply refer to the figure numbers in
% the text per {\it Science\/} style, and include the list of captions at
% the end of the document, coded as ordinary paragraphs as shown in the
% \texttt{scifile.tex} template file.  Your actual figure files should
% be submitted separately.

\end{document}